\documentclass[prd, twocolumn]{revtex4}

\usepackage{graphicx}
\usepackage{amsmath}
\usepackage{amsfonts}
\usepackage{bm}
\usepackage{bbm}
\usepackage{array}
\usepackage{txfonts}

\usepackage{hyperref}

\def\slashchar#1{\setbox0=\hbox{$#1$}     		
   \dimen0=0.25cm                                	
   \setbox1=\hbox{/} \dimen1=\wd1               	
   \ifdim\dimen0>\dimen1                        	
      \rlap{\hbox to \dimen0{\hfil/\hfil}}      	
      #1                                        	
   \else                                        	
      \rlap{\hbox to \dimen1{\hfil$#1$\hfil}}   	
      /                                         	
   \fi}

\renewcommand{\vec}{\boldsymbol}

\newcommand{\rmi}{\mathrm{i}}
\newcommand{\rmd}{\mathrm{d}}
\newcommand{\rme}{\mathrm{e}}
\newcommand{\rmx}{\mathrm{x}}
\newcommand{\rmy}{\mathrm{y}}
\newcommand{\rmz}{\mathrm{z}}

\begin{document}

\title{Dynamics of the Chiral Magnetic Effect in a weak magnetic field}

\author{Harmen J. Warringa}
\email{warringa@th.physik.uni-frankfurt.de}

\affiliation{Frankfurt Institute for Advanced Studies (FIAS), 
Frankfurt am Main, Germany \\
and Institut f\"ur Theoretische Physik, Goethe-Universit\"at 
  Frankfurt am Main, Germany}

\date{\today}

\begin{abstract}
  We investigate the real-time dynamics of the chiral magnetic effect
  in quantum electrodynamics (QED) and quantum chromodynamics
  (QCD). We consider a field configuration of parallel
  (chromo)electric and (chromo)magnetic fields with a weak
  perpendicular electromagnetic magnetic field. The chiral magnetic
  effect induces an electromagnetic current along this perpendicular
  magnetic field, which we will compute using linear response
  theory. We discuss specific results for a homogeneous sudden
  switch-on and a pulsed (chromo)electric field in a static and
  homogeneous (chromo)magnetic field. Our methodology can be easily
  extended to more general situations. The results are useful for
  investigating the chiral magnetic effect with heavy ion collisions
  and with lasers that create strong electromagnetic fields. As a side
  result we obtain the rate of chirality production for massive
  fermions in parallel electric and magnetic fields that are static
  and homogeneous.
\end{abstract}

\maketitle

\section{Introduction}
In quantum electrodynamics (QED) and quantum chromodynamics (QCD)
certain gauge field configurations will induce chirality as follows
from the axial Ward identity \cite{S51, ABJ}. These gauge field
configurations have a net topological charge density, which means that
the contraction of the field strength tensor with its dual is
nonvanishing. An example of such gauge field configuration is one in
which the (chromo)electric and (chromo)magnetic fields are
parallel. For massless particles, chirality is the asymmetry between
the number of particles plus antiparticles with right-handed helicity
and the number of particles plus antiparticles with left-handed
helicity. Here right-handed helicity means that spin and momentum are
parallel, whereas left-handed helicity means they are anti-parallel.
If a system of charged fermions possesses a net chirality, then an
applied (electromagnetic) magnetic field will induce an
electromagnetic current in the direction of the magnetic field
\cite{V80}. This is called the chiral magnetic effect. In magnetic
fields that are large enough to fully polarize the fermions, one can
easily convince oneself that the magnitude of this current is equal to
the chirality times the absolute charge of the fermion \cite{KMW}.

To improve the understanding of the chiral magnetic effect it is
important to obtain the magnitude of the induced electromagnetic
current in different situations. In many studies of the chiral
magnetic effect the chirality is introduced by hand through a chiral
chemical potential $\mu_5$ \cite{V80, mu5, mu5f}. However, in order to get a
full understanding of the chiral magnetic effect, one should include
the dynamics that leads to a net chirality. Such dynamical studies
have been performed in Euclidean space-time using lattice QCD
\cite{lattice} and with analytic methods, both in Euclidean
\cite{NAM09,BDK12} and Minkowski \cite{FKW10} space-time.

\begin{figure}[t]
\includegraphics[scale=1]{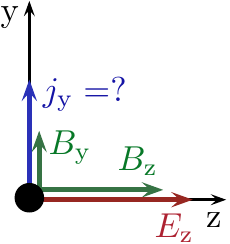}
\caption{Field configuration in which the chiral magnetic effect is
  studied in this article. The electric field along the
  $\rmz$-direction ($E_\rmz$) is homogeneous and can have arbitrary
  time dependence. The magnetic field along the $\rmz$-direction is
  static and homogeneous. The goal is to compute the current density
  $j_\rmy$ along the magnetic field in the $\rmy$-direction
  ($B_\rmy$).
\label{fig:fields}
}
\end{figure}

In this article we will study the real-time dynamics of the chiral
magnetic effect with an analytic method.  We will consider a specific
gauge field configuration of parallel homogeneous (chromo)electric and
(chromo)magnetic fields that are pointing in the
$\rmz$-direction. These parallel fields are the source of the
chirality. The (chromo)magnetic field is assumed to be constant in
time, but the (chromo)electric field can have any time
dependence. Perpendicular to these fields we will consider an
electromagnetic magnetic field with arbitrary time dependence that
points in the $\rmy$-direction.  We have sketched the situation in
Fig.~\ref{fig:fields}. The goal of this article is to obtain the
current density in the $\rmy$-direction, which is due to the chiral
magnetic effect.

The chiral magnetic effect will take place unsuppressed if the fields
have a magnitude that is at least of order of the fermion mass
squared. In the near future, one hopes to create such enormous
electromagnetic fields with high intensity lasers (see
e.g.~\cite{D08}). If one could engineer a field configuration like the
one of Fig.~\ref{fig:fields} with these lasers, one could falsify the
chiral magnetic effect in QED in a controlled environment.  In heavy
ion collisions a similar configuration built out of flux-tubes of
parallel chromoelectric and chromomagnetic fields is induced
\cite{glasma}. Perpendicular to these color fields a large
electromagnetic magnetic field is generated through the currents
created by the colliding charged ions \cite{KMW, SIT}.

In QCD the probability of generating gauge field configurations with
positive and negative topological charge is equal (assuming $\theta =
0$). Hence in heavy ion collisions, the current along the
perpendicular field vanishes on average. However, because of
fluctuations in the initial state \cite{glasma}, and the presence of
sphalerons \cite{M11} the topological charge fluctuates. Hence the
current along the perpendicular magnetic field will fluctuate as well.
That will lead to differences in net electric charge between the two
sides of the reaction plane (the plane in which the beam-axis and the
impact parameter lie) on an event-by-event basis \cite{KKZ}. This
effect can be analyzed experimentally through a charge correlation
study \cite{V04}. Charge correlations have indeed been observed at
both the Relativistic Heavy Ion Collider (RHIC) \cite{STAR} and the
Large Hadron Collider (LHC) \cite{ALICE}. However, currently it is not
clear if these correlations are caused by the chiral magnetic effect
or by another mechanism \cite{alternative}. A possible observation of
the chiral magnetic effect in QCD can teach us about the relevance of
configurations with topological charge.

Theoretically, the field configuration displayed in
Fig.~\ref{fig:fields} is interesting, because when all fields are
static and homogeneous it is possible to obtain a full analytic
expression for the induced current density along the perpendicular
magnetic field \cite{FKW10}.  The current density behaves exactly
according to the qualitative expectations of the chiral magnetic
effect. For example, for a large perpendicular magnetic field the
magnitude of the current was found to be equal to the chirality times
the absolute value of the charge of the fermion. Furthermore, the
current density is suppressed by the fermion mass, and vanishes if
there is no chirality production.

The calculation in Ref.~\cite{FKW10} was build upon a nice argument
based on Lorentz transformations and the particle production rate in
parallel, homogeneous and static fields.  However, this argument
cannot be applied to more general situations that are useful to
investigate. For example, in heavy ion collisions the magnetic field
has a strong time dependence. Furthermore, with high-intensity lasers
one will create inhomogeneous and time-dependent fields. And for
purely academic reasons it is of interest to analyze the real-time
dynamics of the chiral magnetic effect in a spherical symmetric
background gluon field that carries one unit of topological charge,
and is a solution of the Yang-Mills equations.

To obtain the induced current density along the magnetic field in a
general situation one can in principle just solve the Dirac equation
numerically, and construct the current from the solutions.  However,
this might be quite difficult in practice. In order to make progress,
we will start a program to investigate the chiral magnetic effect
using linear response theory. We will expand to first order in the
electromagnetic field that is the source of the perpendicular magnetic
field. In this way we can investigate the chiral magnetic effect for
weak perpendicular magnetic fields that have arbitrary space and time
dependence.  In practical terms we will confirm in this article the
results of Ref.~\cite{FKW10} in the weak magnetic field
limit. Furthermore we will analyze the chiral magnetic effect in
pulsed electric fields.

Most of the calculations in this article will be performed in QED. From
the QED results it is easy to obtain the results for QCD as in
Ref.~\cite{FKW10} and as explained at the end of the introduction.  We
hope that our results will inspire the reader to perform other studies
of the chiral magnetic effect using linear response theory.

This article is organized as follows. In Sec.~\ref{sec:propagator} we
will construct the fermion propagator of QED in a background of
parallel electric and magnetic fields. To support Ans\"atze made in
Ref.~\cite{FKW10}, we compute the electromagnetic current density for
massive fermions analytically in Sec.~\ref{sec:par}. In this section
we will also obtain an analytic expression for the rate of chirality
production for massive fermions in parallel electric and magnetic
fields. In Sec.~\ref{sec:lr} we will then apply linear response theory
to obtain the induced current density along a magnetic field that is
perpendicular to these background fields. We will discuss our
numerical results in Sec.~\ref{sec:results} and show among other
things that we can reproduce the results of Ref.~\cite{FKW10} in the
limit of weak magnetic fields. The conclusions of this work can be
found in Sec.~\ref{sec:conclusions}.

Some details of this work can be found in the appendices. In
Appendix~\ref{app:rel} we give several relations involving the
wavefunctions in a magnetic field. Finally in
Appendix~\ref{app:sol} we discuss the wavefunctions in a sudden
switch-on electric field.

The notational conventions we will use in this article are summarized
below. We also explain the relation between the QED and the QCD setup
below.

\subsection*{Notation}
We will use as a metric $g^{\mu \nu} = \mathrm{diag}(1,-1,-1,-1)$.  We
write $p = p^\mu = (p^0, \vec p)$ to denote a 4-vector.  A component
of a four-vector will be written with either numbers or with an
italic sub- or superscript. For example, $p^z = p^3 = -p_z = -p_3$. We
will use roman subscripts to indicate a component of a 3-vector,
i.e. $\vec p = (p_{\rm x}, p_{\rm y}, p_{\rm z}) =(p^{\rm x}, p^{\rm
  y}, p^{\rm z})$. Hence $p^z = p_{\rm z} = -p_z$. Likewise we write
$\vec x = (x_{\rm x}, x_{\rm y}, x_{\rm z})$ which we most often
simplify to $\vec x = (\rm x, \rm y, \rm z)$.

\subsection*{Relation between QCD and QED setup}
As in Ref.~\cite{FKW10} we consider an Abelianized gluon field
configuration of the form $A^a_\mu = \mathcal{A}_\mu n^a$ in the QCD
setup. Here $n^a$ denotes a direction in the adjoint space. Since the
electromagnetic current density, the object we wish to compute, is
gauge invariant, we can perform a gauge transformation to rotate $n^a$
to the $a=3$ direction. From the covariant derivative $D_\mu
= \partial_\mu - i g \mathcal{A}_\mu t^3$, it then follows that the
red quarks effectively experience an electromagnetic field of
magnitude $q A_\mu = -g \mathcal{A}_\mu /2 $ whereas the green quarks
experience an effective electromagnetic field with size $q A_\mu = g
\mathcal{A}_\mu / 2$. The blue quarks are not affected by the
field. Here $q$ is the charge of the fermion and $g$ is the QCD
coupling constant. The current density along the perpendicular
magnetic field in the QCD setup can thus be obtained from the QED
result, yielding
\begin{multline}
  j_{\rmy, \mathrm{QCD}} = 2 \sum_f j_{\rmy, \mathrm{QED}}
(q E_\rmz =  g \mathcal{E}_\rmz/2,
qB_\rmz =  g \mathcal{B}_\rmz / 2, \\
q = q_f, m = m_f). 
\end{multline}
Here $f$ is a sum over quark flavors, $m$ is the mass, and
($\mathcal{E}_\rmz$)$E_\rmz$ and ($\mathcal{B}_\rmz$)$B_\rmz$ denote
respectively the $\rmz$-component of the (chromo)electric and
(chromo)magnetic field.

There is an important difference between the QED and QCD setup.  In
QED the electromagnetic current has also a component in the
$\rmz$-direction due to acceleration of charges along the electric
field. In the QCD setup however, this contribution to the
electromagnetic current vanishes due to a cancellation from the
contribution of red and green quarks. The electromagnetic current in
the QCD setup has thus only a component in the $\rmy$-direction due to
the chiral magnetic effect.

\section{Fermion propagator in electromagnetic background}
\label{sec:propagator}

The goal of this section is to obtain a convenient expression for the
fermion propagator in a background of electric and magnetic fields
both pointing in the $z$-direction. We will consider a homogeneous
static magnetic field, $\vec B = B_\rmz \vec e_\rmz$. The electric field is
assumed to be homogeneous as well with an arbitrary time-dependence,
$\vec E = E(t) \vec e_\rmz$. The corresponding gauge fields will be
chosen as: $A^0 = 0$ and $\vec A = (0, B_\rmz x, A_{\rmz}(t))$ with
$\lim_{t \rightarrow -\infty} A_{\rmz}(t) = 0$ so that $E(t) =
-\partial_t A_{\rmz}(t)$.

In this section we will first compute the Dirac spinors in this
electromagnetic background. Using these spinors we will construct the
quantum field for fermions, from which we will derive the different
fermion propagators.

\subsection{Dirac spinors}

The spinors $\psi(x)$ can be obtained from the Dirac equation which
reads
\begin{equation}
 \left( \rmi \gamma^\mu D_\mu - m \right) \psi(x) = 0,
\label{eq:diraceq}
\end{equation}
where the covariant derivative $D_\mu = \partial_\mu + \rmi q A_\mu$
with $q$ the electric charge of the fermion. 

To solve the Dirac equation in the background given above we will
apply the Ritus method \cite{R72}.  This method amounts to making the
following ansatz for the particle spinors ($+$) with momentum $\vec p$
and antiparticle spinors ($-$) with momentum $-\vec p$
\begin{eqnarray}
  \psi^+_{\vec p s}(x) &=& F^+_{\vec p}(t) 
  G_{\vec p}(\rmx) \rme^{\rmi p_\rmy \rmy + \rmi p_\rmz \rmz} 
  u_s(\tilde p_+),
 \label{eq:diracsolU}
\\
  \psi^-_{\vec p s}(x) &=& 
F^-_{\vec p}(t) G_{\vec p}(\rmx)
  \rme^{\rmi p_\rmy \rmy + \rmi p_\rmz \rmz} v_s(\tilde p_-).
 \label{eq:diracsolV}
\end{eqnarray}
Here the four-vector $\tilde p^\mu_\pm = (\kappa, 0, \mp \lambda, 0)$,
and $u_s(\tilde p_+)$ and $v_s(\tilde p_-)$ denote the usual particle
and antiparticle spinors of the free Dirac equation with
momentum $\tilde p_\pm$. The free Dirac spinors 
satisfy the following equations
\begin{equation}
 (\slashchar{ \tilde p}_+ - m)u_s(\tilde p_+) = 0,
\;\;\;\;
 (\slashchar{ \tilde p}_- + m)v_s(\tilde p_-) = 0.
\label{eq:FreeDirac}
\end{equation}
Furthermore, $\kappa$ and $\lambda$ are constants that depend on $\vec
p$, and will be determined below. Finally, $F_{\vec p}^\pm(t)$
is a $4 \times 4$ matrix that has to commute with $\gamma^1$ and
$\gamma^2$, and $G_{\vec p}(\rmx)$ a $4 \times 4$ matrix that should
commute with $\gamma^0$ and $\gamma^3$.

By inserting the Ritus ansatz into the Dirac equation it follows that
this ansatz is indeed a general solution if the matrices $F^\pm_{\vec
  p}(t)$ and $G_{\vec p}(\rmx)$ obey the following two equations
\begin{eqnarray}
\left[\rmi \gamma^0 \partial_t - \gamma^3 
( p_\rmz - q A_{\rmz}(t)) \right] 
F_{\vec p}^\pm(t) &=& \pm \kappa F^\pm_{\vec p}(t) \gamma^0, 
\label{eq:eleceq}
\\
\left[\rmi \gamma^1 \partial_{\rmx} - \gamma^2 
( p_\rmy - q B_\rmz \rmx) \right] G_{\vec p}(\rmx) 
&=& \lambda G_{\vec p}(\rmx) \gamma^2.
\label{eq:mageq}
\end{eqnarray} 
We will solve these two equations explicitly below.

By multiplying Eq.~(\ref{eq:FreeDirac}) from the left with
$(\slashchar{\tilde p}_\pm \pm m)$ it follows that $\tilde p_{\pm}^2 =
m^2$, so that $\kappa^2 = \lambda^2 + m^2$.  We will normalize the
free Dirac spinors as in Ref.~\cite{PeskinSchroeder}.  In that
convention the polarization sums, which are the only non-trivial
property of the free spinors required in this article, read
\begin{eqnarray}
\sum_{s=\pm} u_s(p) \bar u_s(p) &=& (\slashchar{p} + m),
\label{eq:spinorrelA}
\\
\sum_{s=\pm} v_s(p) \bar v_s(p) &=& (\slashchar{p} - m).
\label{eq:spinorrelB}
\end{eqnarray}

In the remaining part of this subsection, we will determine the
matrices $F_{\vec p}^\pm(t)$ and $G_{\vec p}(\rmx)$, and the values of
$\kappa$ and $\lambda$ explicitly. Without loss of generality both
$F_{\vec p}^\pm(t)$ and $G_{\vec p}(\rmx)$ can be taken diagonal.

\subsubsection{Computation of \texorpdfstring{$G_{\vec p}(\rmx)$}{Gp(x)}}
From Eq.~(\ref{eq:mageq}) it follows that $G_{\vec p}(\rmx)$ is a
function of $\rmx - p_\rmy / qB_\rmz$.  Therefore let us write the diagonal
elements of $G_{\vec p}(\rmx)$ as $g_{a}(\rmx - p_\rmy / q B_\rmz)$
with $a=1\ldots4$. By multiplying Eq.~(\ref{eq:mageq}) from the left with
$\gamma^2$ one finds that
\begin{eqnarray}
\left(\partial_{\rmx} + q B_\rmz \rmx  \right)
 g_{1}(\rmx) &=& \lambda g_{4}(\rmx),
\label{eq:1ordermagA}
\\
\left(
\partial_\rmx - q B_\rmz \rmx
 \right)  g_{4}(\rmx) &=& - \lambda g_{1}(\rmx).
\label{eq:1ordermagB}
\end{eqnarray}
Since the other two components satisfy the same set of equations,
$g_2(\rmx)$ is proportional to $g_4(\rmx)$ and $g_3(\rmx)$ is
proportional to $g_1(\rmx)$. In order to ensure that $G_{\vec
  p}(\rmx)$ commutes with $\gamma^0$ and $\gamma^3$ as required, we
have to take $g_2(\rmx) = g_4(\rmx)$ and $g_1(\rmx) = g_3(\rmx)$.

By combining Eqs.~(\ref{eq:1ordermagA}) and (\ref{eq:1ordermagB}) one
finds that $g_a(\rmx)$ satisfies the following equation
\begin{equation}
 \left( - \partial_\rmx^2 + q^2 B_\rmz^2 \rmx^2 - s q B_\rmz \right) 
g_a(\rmx) = \lambda^2 g_a(\rmx),
\label{eq:harmomag}
\end{equation}
with $s = 1$ for $a=1,3$ and $s=-1$ for $a=2,4$.  Equation
(\ref{eq:harmomag}) is the eigenvalue equation of a harmonic
oscillator with angular frequency $\vert q B_\rmz \vert$.  The eigenvalues
are given by $\lambda^2 = (2 n + 1) \vert q B_\rmz \vert - s q B_\rmz $ with
quantum number $n = 0, 1, 2, \ldots$.  The corresponding
eigenfunctions $\phi_n(\rmx)$ read
\begin{equation}
 \phi_{n}(\rmx) = 
\frac{1}{\sqrt{2^n n!}} 
\left( \frac{ \vert q B_\rmz \vert}{\pi}
\right)^{1/4} \!\!\!
H_n( \sqrt{\vert q B_\rmz \vert}  \rmx) 
\exp \left( 
- \frac{1}{2} 
\vert q B_\rmz \vert \rmx^2 
\right),
\end{equation}
here $H_n(z)$ denotes a Hermite polynomial of degree $n$.
For convenience we define $\phi_{-1}(\rmx) = 0$.
The eigenfunctions $\phi_n(\rmx)$ are normalized as
\begin{equation}
  \int_{-\infty}^{\infty} 
  \mathrm{d}\rmx\, \phi_{n}(\rmx) \phi_{m}(\rmx) 
  = \delta_{nm},
\end{equation}
and satisfy the completeness relation
\begin{equation}
\sum_{n=0}^{\infty} \phi_{n}(\rmx) \phi_{n}(\rmx') 
  = \delta(\rmx-\rmx').
\label{eq:completeness}
\end{equation}
Let us now write $\lambda^2 = 2 k \vert q B_\rmz \vert$ with $k=0, 1, 2,
\ldots$. The full solution to Eq.~(\ref{eq:harmomag}) can be found by
using that $g_1(\rmx)$ and $g_4(\rmx)$ must have the same value of
$\lambda^2$. This gives for $q B_\rmz > 0$, $g_1(\rmx) = c_1 \phi_k(\rmx)$
and $g_4(\rmx) = c_4 \phi_{k-1}(\rmx)$. Choosing the normalization
such that $c_1 = 1$, it follows from Eq.~(\ref{eq:1ordermagA}) that
$c_4 = c_1 = 1$. In this way one obtains also the sign of $\lambda$,
giving $\lambda = \mathrm{sgn}(q B_\rmz) \sqrt{2 \vert q B_\rmz \vert k}$.  As
long as $q B_\rmz > 0$ one then finds $g_a(\rmx) = \phi_{k}(\rmx)$ for
$a=1,3$ and $g_a(\rmx) = \phi_{k-1}(\rmx)$ for $a=2,4$. If $q B_\rmz < 0$
the solution is $g_a(\rmx) = \phi_{k-1}(\rmx)$ for $a=1,3$ and
$g_a(\rmx) = \phi_{k}(\rmx)$ for $a=2,4$. 

The diagonal matrix $G(\rmx)$ can now be written in the following
compact notation
\begin{equation}
 G_{\vec p}(\rmx) = \sum_{s=\pm} g_{\vec p s}(\rmx) \mathcal{P}_{g}^s
\label{eq:gpmfinal}
\end{equation}
where we introduced $g_{\vec p +}(\rmx) = \phi_k( \rmx -
\frac{p_\rmy}{q B_\rmz})$ and $g_{\vec p-}(\rmx) = \phi_{k-1}( \rmx -
\frac{p_\rmy}{q B_\rmz})$. The projection operators $\mathcal{P}_g^s$ are
given by $\mathcal{P}_{g}^s = (1 + \rmi s \, \mathrm{sgn}(q B_\rmz)
\gamma^1 \gamma^2) / 2$.

\subsubsection{Computation of \texorpdfstring{$F^\pm_{\vec p}(t)$}{Fp(t)}}
Let us write the diagonal elements of the matrix $F_{\vec p}^\pm(t)$
as $f^\pm_{a}(t)$ with $a = 1 \ldots 4$. Multiplying
Eq.~(\ref{eq:eleceq}) from the left with $\gamma^0$ yields
\begin{eqnarray}
\left[ \rmi \partial_t + p_\rmz - q A_{\rmz}(t)  \right]
 f^\pm_{1}(t) &=& \pm \kappa f^\pm_{3}(t),
\label{eq:1orderelA}
\\
\left[ \rmi \partial_t - p_\rmz + q A_{\rmz}(t)  \right]
 f^\pm_{3}(t) &=& \pm \kappa f^\pm_{1}(t).
\label{eq:1orderelB}
\end{eqnarray}
Since the other two components satisfy the same set of equations it
follows that $f^\pm_2(t) \propto f^\pm_3(t)$ and $f^\pm_4(t) \propto
f^\pm_1(t)$. In order to ensure that $F_{\vec p}^\pm(t)$ commutes with
$\gamma^1$ and $\gamma^2$ we have to take $f^\pm_2(t) =
f^\pm_3(t)$ and $f^\pm_4(t) = f^\pm_1(t)$. We can therefore write
\begin{equation}
F_{\vec p}^\pm(t) = \sum_{s=\pm} f^\pm_{\vec p s}(t) \mathcal{P}_f^s,
\end{equation}
where $f^\pm_{\vec p+}(t) = f^\pm_1(t)$, $f^\pm_{\vec p-}(t) =
f^\pm_3(t)$ and the projection operators $\mathcal{P}^s_f = (1 + s
\gamma^3 \gamma^0)/2$.

By taking the complex conjugate of Eqs.~(\ref{eq:1orderelA}) and
(\ref{eq:1orderelB}) it can be seen that $f_{\vec p s}^-(t)$ satisfies
the same two differential equations as $f_{\vec p-s}^+(t)^*$. Hence
both functions are proportional to each other.  Since it is natural to
normalize the particle and antiparticle spinors in the same way, we
take
\begin{equation}
f_{\vec p s}^-(t) = f_{\vec p-s}^+(t)^*.
\label{eq:frel1}
\end{equation}
We will use this relation throughout to express our final results in
terms of particle wavefunctions only.

From Eqs.~(\ref{eq:1orderelA}) and (\ref{eq:1orderelB}) it also
follows that the sum $\vert f_{\vec p+}^\pm(t) \vert^2 + \vert f_{\vec
  p-}^\pm(t) \vert^2$ is independent of time. As it turns out it will
be convenient to normalize this combination as
\begin{equation}
 \vert f_{\vec p+}^\pm(t) \vert^2 + \vert f_{\vec p-}^\pm(t) \vert^2 = 2.
\label{eq:frel2}
\end{equation}
In the next subsection we will check that the normalizations we have
made are consistent, by verifying that the quantum fields satisfy the
canonical anti-commutation relations.

In general the wavefunctions $f^\pm_{\vec ps}(t)$ can only be obtained
numerically. There are only a few cases in which analytic solutions
are known. We will now discuss a few of them.

For vanishing electromagnetic field, $A_{\rmz}(t) = 0$, combining
Eqs.~(\ref{eq:1orderelA}) and (\ref{eq:1orderelB}) gives
\begin{equation}
 \left[ - \partial_t^2 - p_\rmz^2
\right] f^\pm_{\vec p s}(t) = \kappa^2 f^\pm_{\vec p s}(t),
\end{equation}
with $s=\pm$.  The solution of the last equation is a linear
combination of phase factors $\exp(\pm \rmi p_0 t)$, where $p_0 =
\sqrt{\kappa^2 + p_\rmz^2}$. The two different solutions correspond to
particles and antiparticles, hence $f^\pm_s(t) = c^\pm_s \exp(\mp
\rmi p_0 t)$ where $c^\pm_s$ is a normalization constant. From
Eq.~(\ref{eq:1orderelA}) it follows that the ratio of the
normalization constants is $c^\pm_+/c^\pm_- = \kappa / (p_0 \pm
p_\rmz)$. Applying Eq.~(\ref{eq:frel2}) gives $(c^\pm_+)^2 +
(c^\pm_-)^2 = 2$, so that
\begin{equation}
 f^\pm_{\vec p s}(t) = \sqrt{\frac{p_0 \mp s p_\rmz}{p_0} } 
\exp(\mp \rmi p_0 t).
\label{eq:fpmfree}
\end{equation}

If $\kappa=0$, the two differential equations for $f^\pm_{\vec p s}(t)$
decouple. In that case Eqs.~(\ref{eq:1orderelA}) and (\ref{eq:1orderelB})
can be integrated straightforwardly, yielding
\begin{equation}
 \left. f^\pm_{\vec p s}(t) \right \vert_{\kappa =0} 
=
\sqrt{2} \exp
\left[ 
\mp \rmi
\vert p_\rmz \vert t
-
\rmi  s \int_{-\infty}^t \rmd t'\,
q A_\rmz(t')
\right] \theta( \mp s p_{\rmz}).
\end{equation}

In a sudden switch-on electric field the wavefunctions $f^\pm_{\vec p
  s}(t)$ are known analytically \cite{N69}. We review the calculation
in Appendix~\ref{app:sol}.  It is also possible to obtain an analytic
solution in a pulsed field of the form $E_\rmz(t) =
E_\rmz/\cosh^2(t/\tau)$ \cite{NN70}.

\subsection{Quantum field}
The Dirac field in the electromagnetic background is given by
\begin{equation}
 \Psi(x) = 
 \sum_{s=\pm} \sum_{\vec p}
\frac{1}{\sqrt{2 \kappa_{\vec p} }} 
\left[
b^{\phantom{\dagger}}_{\vec p s} 
 \psi^{+}_{\vec p s}(x) 
+ d^{\dagger}_{-\vec p s} 
 \psi^{-}_{\vec p s}(x) 
 \right],
\label{eq:quantumfield}
\end{equation}
where we introduced $\sum_{\vec p} \equiv \sum_{k=0}^{\infty} \int
\frac{\mathrm{d} p_\rmy}{2 \pi} \int \frac{\mathrm{d} p_\rmz}{2 \pi}$
and $\kappa_{\vec p} = \kappa$.  Here $\psi^{\pm}_{\vec p s}(x)$ are
the particle ($+$) and antiparticle ($-$) spinors in the background
field, which are given explicitly in Eqs.~(\ref{eq:diracsolU}) and
(\ref{eq:diracsolV}). The operators $b_{\vec p s}$ and $d_{\vec p s}$
denote respectively the annihilation operators for particles and
antiparticles with momentum $\vec p$ and spin $s$ in a background
magnetic field. The creation operator for antiparticles in
Eq.~(\ref{eq:quantumfield}) has negative momentum, reflecting the fact
that in our notation $\psi^{-}_{\vec p s}(x)$ denotes an antiparticle
spinor with momentum $- \vec p$.

The creation and annihilation operators satisfy the following
anti-commutation relations
\begin{multline}
\{
b^{\phantom{\dagger}}_{\vec p s\phantom{'}},\;
b^\dagger_{\vec p' s'}
\} = 
\{
d^{\phantom{\dagger}}_{\vec p s\phantom{'}},\;
d^\dagger_{\vec p' s'}
\} 
\\
= 
(2\pi)^2
\delta_{k k'} \delta(p_\rmy - p_\rmy') \delta(p_\rmz - p_\rmz') 
 \delta_{ss'}.
\end{multline}
All other anti-commutation relations vanish.

To check that all normalization conditions are consistent we will
verify that the quantum field given in Eq.~(\ref{eq:quantumfield})
satisfies the canonical equal-time anti-commutation relation,
which reads
\begin{equation}
 \{ \Psi_a(t, \boldsymbol x), \Psi_b^{\dagger}(t, \boldsymbol x') \}
 = \delta_{ab} \delta^3(\boldsymbol x - \boldsymbol x').
\label{eq:diracanticommutator}
\end{equation}
Using the explicit expression of the quantum field gives after
using the properties of the creation and annihilation operators
\begin{equation}
 \{ \Psi(t, \vec x), \Psi^{\dagger}(t, \vec x') \}
 =
\sum_{u, s=\pm}  
\sum_{\vec p}
\frac{1}{2 \kappa_{\vec p} } 
 \psi^{u}_{\vec p s}(t, \vec x) 
 \psi^{u}_{\vec p s}(t, \vec x')^\dagger.
\end{equation}
Inserting the explicit solution for the spinors, and
summing over spins by applying Eqs.~(\ref{eq:spinorrelA})
and (\ref{eq:spinorrelB}) yields
\begin{multline}
 \{ \Psi(t, \vec x), \Psi^{\dagger}(t, \vec x') \}
 =
 \sum_{u=\pm} \sum_{\vec p}
\frac{1}{2 \kappa_{\vec p}} 
\rme^{\rmi p_\rmy(\rmy-\rmy') + \rmi p_\rmz(\rmz-\rmz')} 
\\
\times \Bigl \{
u F_{\vec p}^u(t) \gamma^0 F_{\vec p}^u(t)^\dagger
\left[m G_{\vec p}(\rmx) G_{\vec p}(\rmx') - 
\lambda G_{\vec p}(\rmx) \gamma^2 G_{\vec p}(\rmx') \right] 
\\
+
\kappa_{\vec p} F_{\vec p}^u(t) F_{\vec p}^u(t)^\dagger 
G_{\vec p}(\rmx) G_{\vec p}(\rmx') 
\Bigr \}
.
\label{eq:anticomb}
\end{multline}
The last equation can be simplified by inserting the explicit
expression for $F_{\vec p}^\pm(t)$. In this way we find that
\begin{eqnarray}
\frac{1}{2} \sum_{u=\pm} F_{\vec p}^u(t) 
 F_{\vec p}^u(t)^\dagger = \openone_4,
\label{eq:frequirementsA}
\\
\sum_{u=\pm} u F_{\vec p}^u(t) \gamma^0 F_{\vec p}^u(t)^\dagger = 0.
\label{eq:frequirementsB}
\end{eqnarray}
To obtain the last two equations, we have used that by combining
Eqs.~(\ref{eq:frel1}) and (\ref{eq:frel2}) it can be shown that
\begin{eqnarray}
&& \vert f^+_{\vec p s}(t) \vert^2 
+
\vert f^-_{\vec p s}(t) \vert^2 = 2,
\\
&& f^+_{\vec p s}(t) 
f^+_{\vec p-s}(t)^*
=
f^-_{\vec p s}(t) 
f^-_{\vec p-s}(t)^*.
\end{eqnarray}
Inserting Eqs.~(\ref{eq:frequirementsA}) and (\ref{eq:frequirementsB})
into Eq.~(\ref{eq:anticomb}) yields
\begin{equation}
 \{ \Psi(t, \vec x), \Psi^{\dagger}(t, \vec x') \}
 = 
\sum_{\vec p}
\rme^{\rmi p_\rmy(\rmy-\rmy') + \rmi p_\rmz(\rmz-\rmz')} 
G_{\vec p}(\rmx) G_{\vec p}(\rmx').
\label{eq:anticomc}
\end{equation}
By using the explicit expression for $G_{\vec p}(\rmx)$ and applying
the completeness relation Eq.~(\ref{eq:completeness}), the canonical
anticommutation relation, Eq.~(\ref{eq:diracanticommutator}), follows
directly. Hence the normalizations we have chosen are consistent.

\subsection{Propagator}
Let us introduce the following definitions for the two-point
correlation functions
\begin{eqnarray}
S^+_{ab}(x, x') &\equiv& \langle 0 \vert \Psi_a(x) 
\bar \Psi_b(x') \vert 0 \rangle,
\\
 S^-_{ab}(x, x') &\equiv& \langle 0 \vert
 \bar \Psi_b(x') \Psi_a(x) 
\vert 0 \rangle.
\end{eqnarray}
Here $\vert 0 \rangle$ denotes the in-vacuum, which in this article is
the vacuum before the electric field has been switched on. The
different propagators (retarded, advanced, Feynman) can be found by
the appropriate linear combinations of the two-point functions. By
applying Eq.~(\ref{eq:quantumfield}) it follows that the two-point
correlation functions expressed in terms of Dirac spinors read
\begin{equation}
S^\pm_{ab}(x, x') 
=
\sum_{s=\pm} \sum_{\vec p}
\frac{1}{2 \kappa_{\vec p}}
\left[\psi^\pm_{\vec p s}(x)\right]_a 
\left[\psi^\pm_{\vec p s}(x')^\dagger \gamma^0 \right]_b
\label{eq:sdef}
.
\end{equation}
By inserting the explicit expressions for the Dirac spinors, and
summing over spins, the two-point correlation functions become
\begin{multline}
S^\pm(x, x') 
= 
\sum_{\vec p}
\frac{1}{2 \kappa_{\vec p}} 
\rme^{\rmi p_\rmy (\rmy-\rmy') + \rmi p_\rmz(\rmz-\rmz')}
\\
\times
F^\pm_{\vec p}(t) G_{\vec p}(\rmx) ( \slashchar{\tilde p}_\pm \pm m)
\gamma^0 F_{\vec p}^\pm(t')^\dagger \gamma^0 G_{\vec p}(\rmx').
\label{eq:sexpl}
\end{multline}

To evaluate the current density and related quantities, one has to contract a
two-point correlation function with an arbitrary combination of gamma
matrices denoted by $\Gamma$. These quantities can be
expressed in terms of $S^\pm(x, x')$ in the following charge symmetric
way,
\begin{multline}
 \langle 0 \vert \bar \Psi(t, \vec x')
 \Gamma \Psi(t, \vec x) \vert 0 \rangle
= 
-
 \frac{1}{2} \sum_{u=\pm} u\,
\mathrm{tr} \left[ S^u(t, \vec x; t, \vec x') \Gamma \right]
\\
+  \frac{1}{2} \mathrm{tr} \left[ \gamma^0 \Gamma \right] 
\delta(\vec x - \vec x')
,
\label{eq:curcs}
\end{multline}
For the current density $j^{\mu}$ we have $\Gamma = q \gamma^\mu$, for
the chirality $n_5$ we have $\Gamma = \gamma^0 \gamma^5$, and for the
pseudoscalar condensate $\Gamma = \rmi \gamma^5$.  Only for $\Gamma =
\gamma^0$ the second trace in Eq.~(\ref{eq:curcs}) does not vanish.

\section{Current and axial anomaly in parallel 
electric and magnetic field}
\label{sec:par}
We will now compute the induced current density, chirality density and
pseudoscalar condensate in the background of parallel homogeneous
time-dependent electric and static magnetic fields. To compute these
quantities we need to evaluate the trace of $S^\pm(x,x') \Gamma$ as
follows from Eq.~(\ref{eq:curcs}). This trace can be easily performed
using a symbolic manipulation program such as Mathematica. Inserting
the explicit forms of $F_{\vec p}^\pm(t)$ and $G_{\vec p}(\rmx)$ gives
\begin{multline}
\mathrm{tr}\left[ S^\pm(x, x') \gamma^3 \right] 
=
- \tfrac{1}{2} \sum_{\vec p} \sum_{r,s=\pm} 
\rme^{\rmi p_\rmy (\rmy-\rmy') + \rmi p_\rmz(\rmz-\rmz')}
\\
\times s f^\pm_{\vec p s}(t) f^\pm_{\vec p s} (t')^* 
g_{\vec p r}(\rmx) g_{\vec p r}(\rmx'),
\end{multline}
\begin{multline}
\mathrm{tr}\left[ S^\pm(x, x') \gamma^0 \gamma^5 \right]
=
- \frac{1}{2} 
\sum_{\vec p} \sum_{r,s=\pm}
\rme^{\rmi p_\rmy (\rmy-\rmy') + \rmi p_\rmz(\rmz-\rmz')} 
\\
\times
\mathrm{sgn}(q B_\rmz)
s f^\pm_{\vec p s}(t) f^\pm_{\vec p s} (t')^* r 
g_{\vec p r}(\rmx) g_{\vec p r}(\rmx'),
\end{multline}
\begin{multline}
\mathrm{tr}\left[ S^\pm(x, x') \gamma^5 \right]
=
\mp 
\sum_{\vec p} \sum_{r,s=\pm}
\frac{m}{2 \kappa_{\vec p}} 
\rme^{\rmi p_\rmy (\rmy-\rmy') + \rmi p_\rmz(\rmz-\rmz')} 
\\
\times
\mathrm{sgn}(q B_\rmz)
s f^\pm_{\vec p s}(t) f^\pm_{\vec p-s} (t')^* 
 r g_{\vec p r}(\rmx) g_{\vec p r}(\rmx').
\end{multline}

We have to evaluate these correlators at equal time ($t=t'$) with the
$\rmx$ and $\rmy$-components of the direction vectors equal
($\rmx=\rmx'$ and $\rmy=\rmy'$). For reasons we explain below, we will
keep $\rmz$ and $\rmz'$ different and introduce $\Delta \equiv \rmz -
\rmz'$. 

In this situation we can perform the $p_\rmy$ integration using the
orthogonality properties of the functions $g_{\vec p s}(\rmx)$.
This yields
\begin{multline}
-\frac{1}{2} \sum_{u=\pm} u \, 
\mathrm{tr} 
\left[S^u(t, \Delta) \gamma^3\right]
=
\\
\frac{\vert q B_\rmz \vert}{4 \pi}
\sum_{k=0}^{\infty} 
\alpha_k 
\int \frac{\rmd p_\rmz}{2\pi}
\rme^{\rmi p_\rmz \Delta} 
\sum_{s=\pm} s \vert f^+_{\vec p s}(t) \vert^2,
\label{eq:j3delta}
\end{multline}
\begin{multline}
- \frac{1}{2} \sum_{u=\pm} u \, 
\mathrm{tr}\left[ S^u(t, \Delta) \gamma^0 \gamma^5 
\right]
=
\\
\frac{q B_\rmz}{4\pi}
\int \frac{\rmd p_\rmz}{2\pi}
\rme^{\rmi p_\rmz \Delta } 
\sum_{s=\pm}
s \vert f^+_{\vec p s}(t) \vert^2 \Bigr \vert_{k = 0}
,
\label{eq:n5delta}
\end{multline}
\begin{multline}
-\frac{1}{2} 
\sum_{u=\pm} u \,
\mathrm{tr}\left[ S^u(t,\Delta) \gamma^5 \right]
=
\\
\mathrm{sgn}(m)
\frac{q B_\rmz}{4\pi}
\int \frac{\rmd p_\rmz}{2\pi}
\rme^{\rmi p_\rmz \Delta } 
\sum_{s=\pm} s f^+_{\vec p s}(t) f^+_{\vec p-s} (t)^*
\Bigr \vert_{k = 0}
\label{eq:etapdelta}
.
\end{multline}
where $\alpha_k = 1$ for $k=0$ and $\alpha_k = 2$ for $k>0$. We have
made use of Eqs.~(\ref{eq:frel1}) and (\ref{eq:frel2}) to express all
results in terms of particle wavefunctions $f^+_{\vec p s}(t)$.

The quantities we will compute can be obtained from
Eqs.~(\ref{eq:j3delta})-(\ref{eq:etapdelta}) in the limit $\Delta
\rightarrow 0$. Naively putting $\Delta = 0$ will not give a well
defined result, due to the presence of ultraviolet divergences.
Therefore we have to regularize our expressions. For consistency, this
regularization should be performed in a gauge invariant way.

A natural way to achieve this is by using the point-split
regularization \cite{AGP83}.  Instead of evaluating the two-point
functions at $\rmz=\rmz'$ one computes them at $\rmz-\rmz' \equiv
\Delta$ and integrates them over a distribution $h(\Delta)$. This
distribution has to be normalized to unity, and should be sharply
peaked around $\Delta = 0$. We will choose $h(\Delta) = \exp(-\Delta^2
/ 4 \epsilon) / (2 \sqrt{ \pi \epsilon})$ and take the limit $\epsilon
\rightarrow 0$.

In order to maintain gauge invariance, the correlators $S^{\pm}(t,
\Delta)$ have to be augmented by a gauge link link $U$ connecting $z$
with $z'$. This gauge link is for both $S^+(t,\Delta)$ and $S^-(t,
\Delta)$ given by
\begin{equation}
 U(t, \Delta) =
\exp \left[
\rmi q \int_{z'}^{z} \mathrm{d} x^\mu A_\mu(x)
\right] = \exp \left[ -\rmi q A_{\rmz}(t) \Delta \right].
\end{equation}
Summarizing, the full point-split regularization prescription reads
\begin{equation}
\mathrm{tr}_r \left[ S^\pm(t) \Gamma \right]
=
\lim_{\epsilon \rightarrow 0}
\int \rmd \Delta
h(\Delta)
U(t, \Delta) \,
\mathrm{tr}\left[ S^\pm(t; \Delta) \Gamma \right],
\end{equation}
here the subscript $r$ stands for regularized. 

We will now apply this regularization prescription to
Eqs.~(\ref{eq:j3delta})-(\ref{eq:etapdelta}) to evaluate the current
density, chirality density and pseudoscalar condensate.  The integral
over $\Delta$ can be performed exactly giving
\begin{multline}
j_z(t)
=
\lim_{\epsilon \rightarrow 0}
q \frac{\vert q B_\rmz \vert}{2 \pi}
\sum_{k=0}^{\infty} 
\alpha_k 
\int \frac{\rmd p_\rmz}{2\pi}
\rme^{ - \epsilon [p_\rmz - q A_{\rmz}(t)]^2  } 
\\
\times
\left[ \vert f^+_{\vec p +}(t) \vert^2 - 1 \right],
\label{eq:regj3}
\end{multline}
\begin{multline}
n_5(t) 
=
\lim_{\epsilon \rightarrow 0}
\frac{q B_\rmz}{2\pi}
\int \frac{\rmd p_\rmz}{2\pi}
\rme^{ - \epsilon [p_\rmz - q A_{\rmz}(t)]^2  } 
\\
\times
\left[ \vert f^+_{\vec p +}(t) \vert^2_{k = 0} - 1 \right],
\label{eq:regn5} 
\end{multline}
\begin{multline}
\eta(t) =
\lim_{\epsilon \rightarrow 0}
\rmi\,\mathrm{sgn}(m)
\frac{q B_\rmz}{4\pi}
\int \frac{\rmd p_\rmz}{2\pi}
\rme^{ - \epsilon [p_\rmz - q A_{\rmz}(t)]^2  }
\\
\times 
\sum_{s=\pm} s f^+_{\vec p s}(t) f^+_{\vec p-s} (t)^*
\Bigr \vert_{k = 0},
\label{eq:regetap}
\end{multline}
where we used that from Eq.~(\ref{eq:frel2}) it follows that $\sum_s
\vert f_{\vec p s}^+(t) \vert^2 = 2 ( \vert f^+_{\vec p +}(t) \vert^2
- 1)$. In the absence of a regulator and in the limit of vanishing
magnetic field, Eq.~(\ref{eq:regj3}) agrees with the results obtained
in Ref.~\cite{K92}.

We can now verify the axial anomaly relation. 
In the massless limit one can show that $\vert f^+_{\vec p+}(t) \vert^2_{k=0}
- 1 = -\mathrm{sgn}(p_\rmz)$. We can now perform the $p_\rmz$ integral
in Eq.~(\ref{eq:regn5}) in the limit $\epsilon \rightarrow 0$ giving
\begin{equation}
n_5(t) = -\frac{q^2}{2\pi^2} B_\rmz A_{\rmz}(t).
\end{equation}
Taking the derivative with respect to time in the last equation gives
the axial anomaly relation for massless particles in parallel electric
and magnetic fields
\begin{equation}
\frac{\rmd n_5(t)}{\rmd t} = \frac{q^2}{2\pi^2} B_\rmz E_\rmz(t).
\end{equation}

If the fermions are massive the anomaly relation contains an
additional term proportional to the pseudoscalar condensate. To obtain
the anomaly relation in this case we will perform the time
derivative on the chirality given in Eq.~(\ref{eq:regn5})
explicitly. The limiting procedure $\epsilon \rightarrow 0$ is
equivalent to performing the integration over $p_\rmz$ in an interval
symmetric around $p_\rmz = q A_\rmz(t)$. Therefore
\begin{equation}
n_5(t) 
=
\lim_{\Lambda \rightarrow \infty}
\frac{q B_\rmz}{2\pi}
\int^{\Lambda + q A_\rmz(t)}_{-\Lambda + q A_\rmz(t)} \frac{\rmd p_\rmz}{2\pi}
\left[ \vert f^+_{\vec p +}(t) \vert^2_{k = 0} - 1 \right].
\end{equation}
The time derivative of $n_5(t)$ contains a part arising from the
integration boundaries and a part from the derivative on the wave-functions.
It follows directly from Eq.~(\ref{eq:1orderelA}) that 
\begin{equation}
 \partial_t \vert f_{\vec p+}(t) \vert^2 = 
  \rmi \kappa_{\vec p} \sum_{s=\pm} s f_{\vec p s}^+(t) f_{\vec p -s}^+(t)^*.
\end{equation}
Also we will use that $\kappa_{\vec p} \vert_{k=0} = \vert m \vert$. 
Furthermore, for large $p_\rmz$ one can neglect $qA_\rmz(t)$,
and from Eq.~(\ref{eq:fpmfree}) it can be shown that
$\lim_{p_\rmz \rightarrow \infty} \vert f_{\vec p+}(t)\vert^2
= 0$ and  $\lim_{p_\rmz \rightarrow -\infty} \vert f_{\vec p+}(t)\vert^2
= 2$. By using Eq.~(\ref{eq:regetap}) and applying the time derivative to
the integration boundaries it follows that
\begin{equation}
\frac{\rmd n_5(t)}{\rmd t} = 2 m \eta(t) 
+ \frac{q^2}{2\pi^2} B_\rmz E_\rmz(t).
\label{eq:anomalymass}
\end{equation}
which is exactly the anomaly relation in the presence of mass.

Let us now consider a sudden-switch on electric field of the form
$E(t) = E_\rmz \theta(t)$. We discuss the functions $f_{\vec p s}^\pm(t)$ in
this field in Appendix \ref{app:sol}.  We can only compute the induced
current density and the chirality density analytically in the large
$t$ limit. Applying Eq.~(\ref{eq:ilimit}) we find that for large $t$
\begin{eqnarray}
\frac{\rmd j_z(t)}{\rmd t}
 & =& q \frac{\vert q B_\rmz \vert q E_\rmz}{2 \pi^2}
\rme^{-\frac{\pi m^2}{\vert q E_\rmz \vert}}
\sum_{k=0}^{\infty} 
\alpha_k 
\rme^{-  2 \pi k \vert \frac{ q B_\rmz}{q E_\rmz } \vert } 
\nonumber
 \\
&=&
q \frac{\vert q B_\rmz \vert q E_\rmz}{2 \pi^2}
\coth ( \vert \tfrac{B_\rmz}{E_\rmz } \vert \pi )
\rme^{-\frac{\pi m^2}{\vert q E_\rmz \vert}},
\label{eq:jzpar}
\\
\frac{\rmd n_5(t)}{\rmd t}
 &=&
\frac{q^2 E_\rmz B_\rmz}{2 \pi^2}
\rme^{-\frac{\pi m^2}{\vert q E_\rmz \vert}}.
\label{eq:n5par}
\end{eqnarray}
Since these results are obtained for large $t$ in a sudden-switch on
electric field, they are exact in a constant electric field.

The result Eq.~(\ref{eq:jzpar}) was derived analytically in a
different way in Ref.~\cite{GG08}. It was also proved to be correct
numerically in Ref.~\cite{T09}. The result is easy to understand
starting from the production rate of fermion antifermion pairs in
parallel homogeneous electric and magnetic fields \cite{FKW10}. That
rate per unit volume equals \cite{N69, BT70} (see also \cite{D04,
 KP06, CM08})
\begin{equation}
 \Gamma = \frac{q^2 E_\rmz B_\rmz}{4\pi^2} 
  \coth \left(\frac{B_\rmz}{E_\rmz} \pi \right) 
 \exp \left( - \frac{m^2 \pi}{\vert q E_\rmz\vert} 
\right). 
\label{eq:gammapar}
\end{equation}
For $B_\rmz = 0$ this rate reduces to the pair production rate in a
homogeneous electric field, that was first obtained by Schwinger
\cite{S51}.  The production of pairs gives rise to a homogeneous
current density that has to point in the $\rmz$-direction because of
symmetry reasons. The particles are accelerated continuously by the
electric field. Therefore, at some point they will reach (almost) the
speed of light.  Hence, every time a pair is created the current will
eventually grow by twice the charge of the fermion. So therefore the
rate of change of the current density is given by $\partial_t j_\rmz =
2 q \Gamma \mathrm{sgn}(qE_\rmz)$. Inserting Eq.~(\ref{eq:gammapar})
we see that we exactly recover Eq.~(\ref{eq:jzpar}). One can also use
this argument in the opposite order, in order to derive the pair
production rate from the calculation of the current density.

The result Eq.~(\ref{eq:n5par}) generalizes the well-known production
rate of chirality in parallel electric and magnetic fields for
massless fermions to massive fermions. We are unaware of an earlier
derivation of this result. The mass suppresses the production of
chirality. By combining Eq.~(\ref{eq:n5par}) with
Eq.~(\ref{eq:anomalymass}) we find that the pseudoscalar condensate in
static, homogenous and parallel electric and magnetic fields equals
\begin{equation}
 \eta(t) = \frac{q^2 E_\rmz B_\rmz}{4 \pi^2 m} 
\left( \rme^{-\frac{\pi m^2}{\vert q E_\rmz \vert}} - 1 \right).
\end{equation}

\section{Linear response to magnetic field}
\label{sec:lr}
In the previous section we have considered a time-dependent electric
field and a constant magnetic field that were both pointing in the
$\rmz$-direction. To this field configuration we will now add a
time-dependent magnetic field in the $\rmy$-direction, denoted by
$B_\rmy(t)$.  This magnetic field will be accompanied by a
perpendicular electric field as can be seen from Faraday's law, $\vec
\nabla \times \vec E = -\partial \vec B(t) / \partial t$.  The
additional magnetic field will induce a current density in the
$\rmy$-direction. In this section we will compute this current density
to first order in $B_\rmy(t)$ using linear response theory.  In the
next section we will use this result to study the chiral magnetic
effect.

Let us write the full electromagnetic field as $A^\mu(x) = \bar
A^\mu(x) + \tilde A^\mu(x)$. Here $\bar A^\mu(x)$ denotes the
background field, consisting of the electric and magnetic fields
pointing in the $\rmz$-direction. The field $\tilde A^{\mu}(x)$
denotes the perturbation on this background, which in this case is the
magnetic field in the $\rmy$-direction with its corresponding
perpendicular electric field.

From linear response theory it follows that to first order in $\tilde
A_\mu(x)$ the induced current density in the electromagnetic field
$A^\mu(x)$ equals $j^\mu(x) = j^\mu_A(x) = j^\mu_{\bar A}(x) + \delta
j_{\bar A, \tilde A} ^\mu(x)$ where
\begin{equation}
\delta j^\mu_{\bar A, \tilde A}(x) =
 \int \rmd^4 x' \, \Pi^{\mu \nu}_{\mathrm{R}}(x, x') \tilde
A_\nu(x').
\end{equation}
Here the retarded current-current correlator (or equivalently photon
polarization tensor) in the background field $\bar A_\mu(x)$ is given
by $\Pi^{\mu \nu}_{\mathrm{R}}(x, x') = \Pi^{\mu \nu}_{\mathrm{C}}(x,
x') \theta(t-t')$ with
\begin{equation}
\Pi^{\mu \nu}_{\mathrm{C}}(x, x') = - \rmi q^2 \langle 0 \vert 
\left [ \bar \Psi(x) \gamma^{\mu} \Psi(x),
\bar \Psi(x') \gamma^{\nu} \Psi(x') \right] \vert 0 \rangle
.
\end{equation}
Using Eqs.~(\ref{eq:quantumfield}) and (\ref{eq:sdef}) we can express
$\Pi^{\mu \nu}_{\mathrm{C}}(x, x')$ as
\begin{equation}
\Pi^{\mu \nu}_{\mathrm{C}}(x, x') = 
-\rmi q^2 \sum_{u=\pm} u\, \mathrm{tr}
\left[ \gamma^\mu S^u(x, x') \gamma^\nu S^{-u}(x', x) 
\right],
\label{eq:pimunuc}
\end{equation}
where $S^\pm( x, x')$ is the two-point correlation function in the
background field, given explicitly in Eq.~(\ref{eq:sexpl}).

Since the background electric and magnetic fields are both pointing in
the $\rmz$-direction, they cannot solely induce a current in the
$\rmy$-direction. As a result $j^{\rmy}_{\bar A}(t) = 0$.  Hence the
induced current density in the $\rmy$-direction can only arise from
the perturbation and is therefore of the following form
\begin{equation}
j_{\rmy}(t) =
\int_{-\infty}^{t}
\rmd t'
 H(t, t') B_{\rmy}(t').
\label{eq:jcm}
\end{equation}
where $H(t,t')$ can be obtained from $\Pi^{\mu \nu}_{\mathrm{C}}(x,
x')$ as we will explain in more detail below. The photon polarization
tensor in an electric plus magnetic background has been studied by
other authors before in different contexts
\cite{PhotonPol}. Furthermore, the photon polarization tensor in a
purely magnetic background has been studied in detail in several works
\cite{PhotonPolMag}, for recent analyses and applications we refer to
\cite{mu5, PhotonPolMagRecent}.

In the following subsections we will compute the function $H(t, t')$
in two cases, labeled by A and B. In case A we will take the
only non-vanishing component of the perturbation field to be $\tilde
A_\rmz(x) = -B_{\rm y}(t) \rmx$. In case B the only non-vanishing
component is chosen as $\tilde A_\rmx(x) = B_{\rm y}(t) \rmz$. These two
cases lead to the same magnetic field in the $\rmy$-direction,
$B_{\rmy}(t)$, but give rise to different perpendicular electric
fields. The only non-vanishing component of the additional electric
field is in the first case $E_\rmz = \partial_t B_{\rmy}(t) \rmx$, and
in the second case $E_\rmx = -\partial_t B_{\rmy}(t) \rmz$. By taking
the average of the two cases, one obtains a more symmetric electric
field, which is circular in the $\rmx$-$\rmz$ plane.

If $B_{\rmy}(t)$ is constant in time, the perpendicular electric field
vanishes and the two cases are gauge equivalent. However, we do
not know how to implement a constant magnetic field exactly in a
practical numerical calculation. In the numerical evaluation of
Eq.~(\ref{eq:jcm}) one might replace the lower integration bound
by a finite time $t_a$.  But then one effectively deals with a sudden
switch-on perpendicular field of the following form
\begin{equation}
 B_\rmy(t) = \left \{ \begin{array}{cl}
0 & \mathrm{if}\; t < t_a, \\
B_\rmy & \mathrm{if}\; t \geq t_a .
\end{array}
\right.
\end{equation}
This sudden switch-on leads to large perpendicular electric fields,
and as a result the current density in case A and B can be different. To
instead implement a magnetic field that is effectively constant in
time we have to switch it on slowly. In our numerical calculations we
will take a magnetic field of the following form
\begin{equation}
 B_\rmy(t) = \left \{ \begin{array}{cl}
0 & \mathrm{if}\; t < t_a, \\
B_\rmy(t-t_a)/(t_b-t_a) & \mathrm{if}\;  t_a \leq t \leq t_b, \\
B_\rmy & \mathrm{if}\; t > t_b.
\end{array}
\right.
\label{eq:slowb}
\end{equation}
If we choose $t_b - t_a$ large enough and $t_b$ small enough before
the important physical effects happen, the magnetic field is
effectively constant. In that situation the perpendicular electric fields
are small and the induced current density in case A and B should approximately
have the same magnitude. We will use this feature to test our
methodology.

\subsection{\texorpdfstring{$\tilde A_\rmz(x) = -B_{\rmy}(t) \rmx $}{Az(x) = -By(t) x}}
In the case that $\tilde A_\rmz(x) = -B_{\rmy}(t) \rmx $ we obtain
\begin{equation}
  H(t, t') = \int \mathrm{d}^3 \vec x' \, \mathrm{x}'
 \, \Pi^{23}_{\mathrm{C}}(x, x').
\end{equation}
After inserting the explicit expression for $S^\pm(x, x')$, taking the
trace, and integration over $\rmy$, $\rmz$, $p_\rmy'$ and $p_\rmz'$ we
find
\begin{multline}
  H(t, t') = -q^2
\sum_{\vec p}
\sum_{k'=0}^{\infty}
\int_{-\infty}^{\infty} \rmd \rmx'\, \rmx'
\\
\times
\left [
\frac{\lambda_{\vec p}}{\kappa_{\vec p}}
V_{\vec p, \vec p'}(t,t') W_{\vec p, \vec p'}(\rmx, \rmx')
\right ]_{p_\rmy'=p_\rmy, p_\rmz'=p_\rmz},
\label{eq:hvw1}
\end{multline}
where the functions $V$ and $W$ are given by 
\begin{eqnarray}
V_{\vec p, \vec p'}(t,t') &=& 
\mathrm{Im}
\sum_{s=\pm} s
f^+_{\vec ps}(t)
f^+_{\vec p's}(t)
f^+_{\vec p-s}(t')^*
f^+_{\vec p's}(t')^*,
\\
W_{\vec p, \vec p'}(\rmx, \rmx')
&=&
\sum_{s=\pm} 
g_{\vec p-s}(\rmx)
g_{\vec p's}(\rmx)
g_{\vec ps}(\rmx')
g_{\vec p's}(\rmx').
\end{eqnarray}

The expression for $H(t,t')$ can be simplified by performing the
integration over $\rmx'$ followed by integration over $p_\rmy$. 
Using the relations from Appendix~\ref{app:rel} it can be shown that
\begin{multline}
\int_{-\infty}^{\infty} \frac{ \rmd p_\rmy}{2 \pi }
\int_{-\infty}^{\infty} \rmd \rmx' \, \rmx' \, 
W_{\vec p, \vec p'}(\rmx,\rmx') \Bigr \vert_{p_\rmy'=p_\rmy}
\\
= 
\frac{\sqrt{\vert q B_\rmz \vert}}{2 \pi \sqrt{2}}
\sqrt{k} 
\left[
2 
\delta_{k, k'}
-
\delta_{k-1, k'}
-
\delta_{k+1, k'}
\right].
\label{eq:intw}
\end{multline}
Inserting Eq.~(\ref{eq:intw}) into Eq.~(\ref{eq:hvw1}) yields
\begin{multline}
H(t, t') = 
\frac{q^2}
{2\pi}
\sum_{k=1}^{\infty}
\int_{-\infty}^{\infty} \frac{\mathrm{d} p_\rmz}{2\pi}
\omega_k
\times
\\
\left[
2 V_{k,k}(t, t')
-
V_{k,k-1}(t, t')
-
V_{k, k+1}(t, t')
\right]_{p_\rmz = p_\rmz'},
\label{eq:H1a}
\end{multline}
where 
\begin{equation}
  \omega_k = 
\frac{\vert q B_\rmz \vert k}{\sqrt{2 \vert q B_\rmz \vert k + m^2}}.
\end{equation}
To speed up the numerical computation it is convenient to take
the wavefunctions with the same momenta together in the integrand.
For this reason we rewrite Eq.~(\ref{eq:H1a}) into
\begin{multline}
H(t, t') = 
-\frac{q^2}
{2\pi}
\int_{-\infty}^{\infty} \frac{\mathrm{d} p_\rmz}{2\pi}
\omega_1 V_{1,0}(t, t') +
\frac{q^2}
{2\pi} \sum_{k=1}^{\infty}
\int_{-\infty}^{\infty} \frac{\mathrm{d} p_\rmz}{2\pi} 
\times
\\
\left[
2 \omega_k V_{k,k}(t, t')
-
\omega_{k+1}V_{k+1,k}(t, t')
-
\omega_k V_{k, k+1}(t, t')
\right]_{p_\rmz = p_\rmz'}.
\label{eq:H1}
\end{multline}

\subsection{\texorpdfstring{$\tilde A_\rmx(x) = B_{\rmy}(t) \rmz $}{Ax = By(t)}}
If $\tilde A_\rmx(x) = B_{\rmy}(t) \rmz $ we obtain
\begin{equation}
  H(t, t') = -\int \mathrm{d}^3 \vec x' \, \rmz'
 \, \Pi^{21}_{\mathrm{C}}(x, x').
\end{equation}
Inserting the explicit expression for the two-point function and
performing the trace, we find that
\begin{multline}
 \Pi^{21}_{\mathrm{C}}(x, x') = 
 \rmi \frac{q^2}{2} \sum_{\vec p, \vec p'} 
\rme^{
\rmi(p_\rmy - p_\rmy')(\rmy - \rmy')
+
\rmi(p_\rmz - p_\rmz')(\rmz - \rmz')}
\times
\\
\left \{
W^A_{\vec p, \vec p'}(\rmx, \rmx') 
\left[
V^A_{\vec p, \vec p'}(t, t')
+ \frac{m^2}{\kappa_{\vec p} \kappa_{\vec p'}}
V^B_{\vec p, \vec p'}(t, t')
 \right]
\right.
\\
+ \left.
\frac{\lambda_{\vec p} \lambda_{\vec p'}}
{\kappa_{\vec p} \kappa_{\vec p'}}
W^B_{\vec p, \vec p'}(\rmx, \rmx') 
V^B_{\vec p, \vec p'}(t, t')
\right \},
\end{multline}
where we have defined the following functions
\begin{eqnarray}
V^A_{\vec p, \vec p'}(t,t')  &=& 
\mathrm{Im}
\sum_{s=\pm} 
f^+_{\vec ps}(t)
f^+_{\vec p's}(t)
f^+_{\vec ps}(t')^*
f^+_{\vec p's}(t')^*,
\\
V^B_{\vec p, \vec p'}(t,t') &= 
&
\mathrm{Im}
\sum_{s=\pm} 
f^+_{\vec ps}(t)
f^+_{\vec p's}(t)
f^+_{\vec p-s}(t')^*
f^+_{\vec p'-s}(t')^*,
\\
W^A_{\vec p, \vec p'}(\rmx, \rmx')
&=&
\sum_{s=\pm} s
g_{\vec p s}(\rmx)
g_{\vec p'-s}(\rmx)
g_{\vec ps}(\rmx')
g_{\vec p'-s}(\rmx'),
\\
W^B_{\vec p, \vec p'}(\rmx, \rmx') 
&=&
\sum_{s=\pm} s
g_{\vec ps}(\rmx)
g_{\vec p'-s}(\rmx)
g_{\vec p-s}(\rmx')
g_{\vec p's}(\rmx').
\end{eqnarray}

To simplify $H(t,t')$ we can make use of the following relation
\begin{multline}
\int \frac{\rmd p_\rmz}{2\pi}
\int \frac{\rmd p'_\rmz}{2\pi}
\int \rmd \rmz' \,
\rmz' \rme^{\rmi(p_\rmz-p_\rmz')(\rmz - \rmz')}
f(p_\rmz, p_\rmz') =
\\
-\rmi 
\int \frac{\rmd \bar p_\rmz}{2\pi}
\left[
\frac{\partial}{\partial h} 
\rme^{\rmi \rmz h}
f(\bar p_\rmz+h/2, \bar p_\rmz - h/2)
\right]_{h=0}.
\end{multline}
Furthermore, using the relations from Appendix~\ref{app:rel} 
it follows that
\begin{multline}
\int_{-\infty}^{\infty} \frac{ \rmd p_\rmy}{2 \pi }
\int_{-\infty}^{\infty} \rmd \rmx' \, 
W^A_{\vec p, \vec p'}(\rmx,\rmx') \Bigr \vert_{p_\rmy'=p_\rmy}
\\
= 
\frac{\vert q B_\rmz \vert}{2 \pi}
\left(
\delta_{k+1, k'}
-
\delta_{k-1, k'} \right),
\label{eq:intWA}
\end{multline}
\begin{equation}
\int_{-\infty}^{\infty} \frac{ \rmd p_\rmy}{2 \pi }
\int_{-\infty}^{\infty} \rmd \rmx' \, 
W^B_{\vec p, \vec p'}(\rmx,\rmx') \Bigr \vert_{p_\rmy'=p_\rmy}
= 0.
\label{eq:intWB}
\end{equation}
Using the last three equations above we find
\begin{multline}
 H(t, t')
= 
- \frac{q^2 \vert q B_\rmz \vert}{2 \pi}
\sum_{k=0}^{\infty}
\int \frac{\rmd \bar p_\rmz}{2\pi}
\frac{\partial}{\partial h}
\Bigl[
V^A_{k, k+1}(t, t') 
\\
+ \frac{m^2}{\kappa_{k} \kappa_{k+1}} V^B_{k, k+1}(t, t')
\Bigr],
\label{eq:H2}
\end{multline}
where $p_\rmz=\bar p_\rmz + h/2$ and $p_\rmz' = \bar p_\rmz - h/2$.

\subsection{Numerical procedure}
We now will discuss the details of the numerical evaluation of the
current density along the perpendicular magnetic field.  Firstly, we
obtain the wavefunctions $f^+_{\vec p s}(t)$ numerically through
solving Eqs.~(\ref{eq:1orderelA}) and (\ref{eq:1orderelB}) using a
Runge-Kutta method implemented in Matlab. We make sure that we obtain
$f^+_{\vec p s}(t)$ at equally spaced time steps. The next step is to
construct the integrand of Eqs.~(\ref{eq:H1}) and (\ref{eq:H2}) for
different $t'$. The derivative in the integrand of Eq.~(\ref{eq:H2})
is computed using finite differences. We then perform the $t'$
integration in Eq.~(\ref{eq:jcm}) using the trapezoidal rule with a
lower integration cutoff. Thereafter we perform the $p_\rmz$ integral,
in an interval symmetric around $p_\rmz = 0$, using the trapezoidal
rule. The upper and lower cutoff are taken so large that varying them
does not change the results. The last step is to perform the sum over
$k$. The results are dominated by the small $k$ values, therefore we
sum over $k$ until we reach convergence. Typically one only has to sum
over a few values of $k$ to obtain an accurate answer.

\section{Results}
\label{sec:results}
We will now study the current density generated by the chiral magnetic
effect in parallel electric and magnetic fields with a perpendicular
magnetic field as in Fig.~1.  We will compute this current density
numerically using the linear response relation
Eq.~(\ref{eq:jcm}). Since the calculation is based on linear response,
our results will be valid for perpendicular magnetic fields small
compared to the parallel electric and magnetic fields.

The full electromagnetic current density has two components. Firstly,
it has a component along the perpendicular magnetic field due to the
chiral magnetic effect. Secondly, it has a component in the
longitudinal direction along the electric and magnetic field due to
pair production. This component does not vanish for weak
perpendicular magnetic fields, and can to first order be computed
using Eq.~(\ref{eq:regj3}). As explained in the introduction the
longitudinal component vanishes in the QCD setup due to a cancellation
of the contribution of red and green quarks.

The electromagnetic current density will generate fields themselves which can
modify the dynamics. This backreaction can be safely neglected as long
as the fields induced by the currents stay small compared to the
background fields.  Such regime can always be reached by considering
times short after the switch-on of the background fields. In this
article we will not consider this backreaction and leave its study for
future work.

We will present results for the chiral magnetic effect in a sudden
switch-on electric field and a pulsed electric field below. We
consider a perpendicular magnetic field that is effectively constant
in time as in Eq.~(\ref{eq:slowb}).  The formalism we have developed
in this article allows one to analyze the chiral magnetic effect in
other settings as well.

We have performed all numerical calculations in an effectively
constant magnetic field using both gauge field choice A and B. These
choices are approximately gauge equivalent, and the accuracy of the
approximation can be improved by switching on the effectively constant
magnetic field slower. The calculations performed using gauge field
choice A and B are independent and hence can be used to test our
methodology. In the numerical calculations in an effectively constant
magnetic field we have found excellent agreement between the results
obtained with choice A and B.

\subsection{Sudden switch-on electric field}
Here we will consider the chiral magnetic effect in a sudden-switch on
electric field of the following form $E(t) = E_\rmz \theta(t)$. The
corresponding gauge field reads $A_\rmz(t) = - E_\rmz \theta(t)t$.
The perpendicular magnetic field is taken to be effectively constant.
In Ref.~\cite{FKW10} the current density in the $\rmy$-direction was
computed exactly for $t\gg 0$. For small $B_\rmy$ and $t \gg 0$ 
the rate of current density generation equals~\cite{FKW10}
\begin{equation}
 \partial_t j_\rmy = \frac{q^2 B_\rmy}{2\pi^2} 
\frac{\vert q E_\rmz \vert B_\rmz^2}{B_\rmz^2 + E_\rmz^2}
\coth \left( \frac{B_\rmz}{E_\rmz} \pi \right)
\exp \left(- \frac{m^2 \pi}{\vert q E_\rmz \vert} \right).
\label{eq:djdt}
\end{equation}

\begin{figure}[ht]
\includegraphics[scale=1]{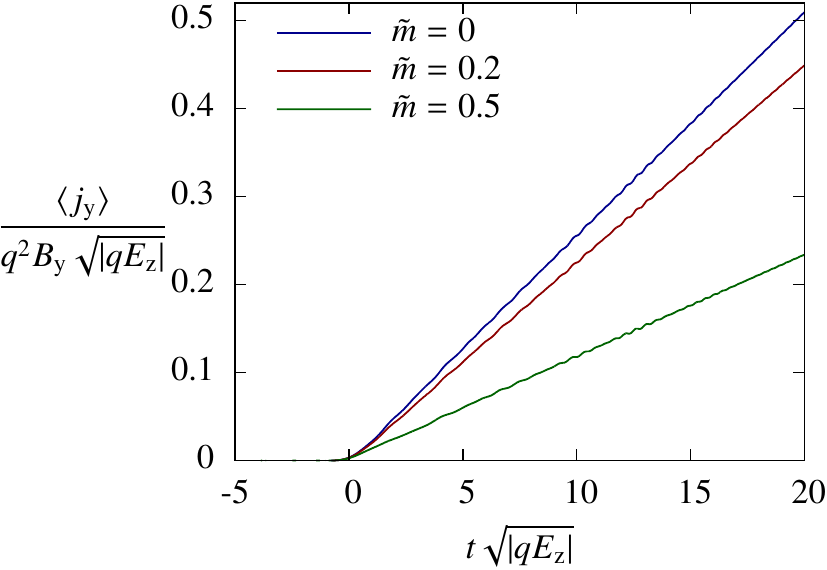}
\caption{Running average of the current density generated by the chiral magnetic
effect in an electric field suddenly switched on at $t=0$. Here $B_\rmz/E_\rmz = 1$.
\label{fig:j_sso}
}
\end{figure}

In order to cancel the rapid oscillations in the current density
arising from the sudden-switch on of the electric field we will
investigate the running average of the current density, here defined as
\begin{equation}
  \langle j_\rmy(t) \rangle = 
  \int_{t-c/\sqrt{\vert qE_\rmz \vert}}^{t+c/\sqrt{\vert qE_\rmz \vert}} 
  \rmd t' \, j_\rmy(t').
\end{equation}
We display the results of the numerical computation of the running
average of the current density using $c=1$ in Fig.~\ref{fig:j_sso} for
different values of $\tilde m = m / \sqrt{\vert q E_\rmz \vert}$ and
$B_{\rmz}/E_{\rmz} = 1$.  The linear response calculation shows that
after the switch-on the current quickly grows linear with time. A
fermion mass suppresses the production of chirality as can be seen
from Eq.~(\ref{eq:n5par}). This explains why the current density is
smaller for particles with a larger mass.

\begin{figure}[t]
\includegraphics[scale=1]{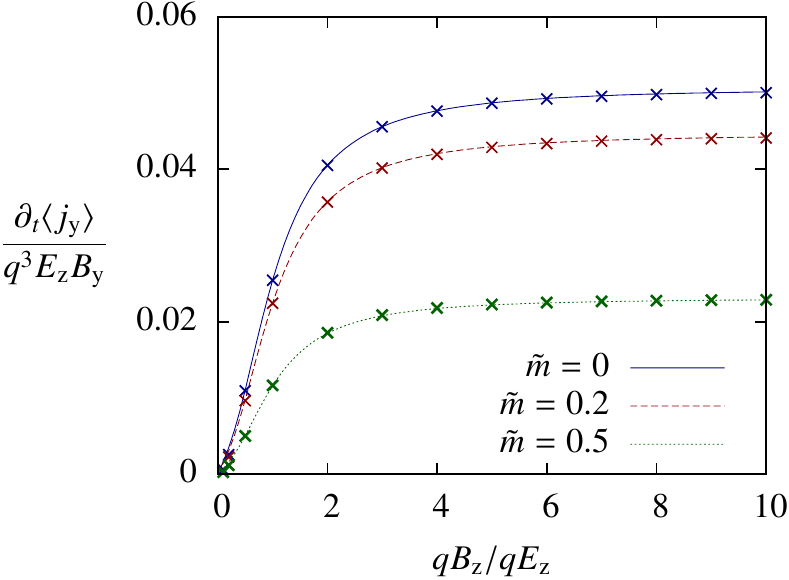}
\caption{Rate of current density generation due to the chiral
magnetic effect at late
times in a sudden-switch on electric field as a function of $B_\rmz/E_\rmz$ 
for different masses. The lines
denote the small $B_\rmy$ limit of the exact result; points indicate the results
of the numerical calculations using linear response.
\label{fig:djdt_sso}
}
\end{figure}

In Fig.~\ref{fig:djdt_sso} we compare our numerical results for the
rate of current density generation to the small $B_\rmy$ limit of the exact
result, given in Eq.~(\ref{eq:djdt}). It can be seen that we find
excellent agreement between the results obtained using linear response
and the small $B_\rmy$ limit of the exact result. Thus our linear
response approach has passed a critical test. It implies that the
study of the dynamics of the chiral magnetic effect using linear
response can be performed successfully. Alternatively, our results can
be seen as an independent verification of the results obtained in
Ref.~\cite{FKW10}.

It can be seen that the rate of current density generation increases if
$B_\rmz$ is enlarged. This is natural, since the amount of chirality
production is increased. But at the same time enlarging $B_\rmz$
decreases the degree of polarization of the fermions in the
$\rmy$-direction. The combination of these two effects results in the
saturation of the rate of current generation for large $B_\rmz$.

\subsection{Pulsed electric field}
We will now study the chiral magnetic effect in a pulsed electric
field that has the form $E_\rmz(t) = E_\rmz / \cosh^2(t/\tau)$.  The
corresponding gauge field reads $A_\rmz(t) = - E_\rmz [1 +
\tanh(t/\tau)] \tau$. The perpendicular magnetic field is taken
to be effectively constant.

\begin{figure}[t]
\includegraphics[scale=1]{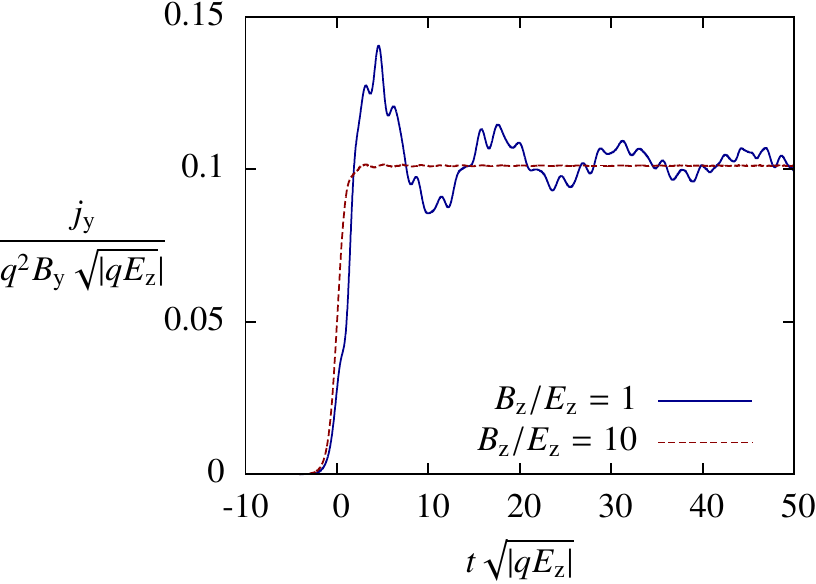}
\caption{Current density due to the chiral magnetic effect
in a pulsed electric field, as a function
of time. Here $\tau = 1/\sqrt{\vert qE_\rmz \vert}$ and $\tilde m = 0$.
\label{fig:jp_m0}
}
\end{figure}

\begin{figure}[t]
\includegraphics[scale=1]{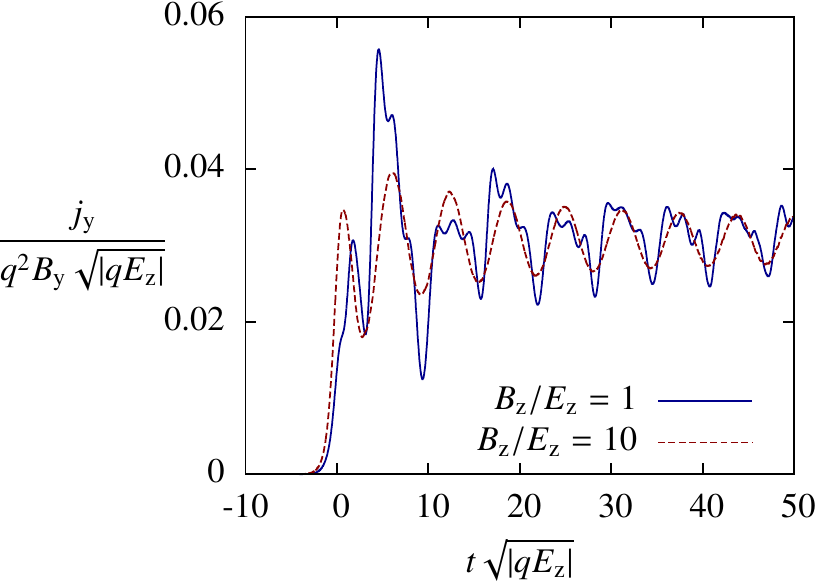}
\caption{Same as in Fig.~\ref{fig:jp_m0} but now for $\tilde m = 0.5$.
\label{fig:jp_m05}
}
\end{figure}

In Figs.~\ref{fig:jp_m0} and \ref{fig:jp_m05} we display the current
density generated by the chiral magnetic effect as a function of time
for $\tau = 1/\sqrt{\vert qE_\rmz \vert}$ and respectively $\tilde m =
m/\sqrt{ \vert q E_\rmz \vert} = 0$ and $\tilde m = 0.5$. It can be
seen that the current density rises quickly around $t=0$. This is
because only then an electric field is present so that chirality will
be produced. For large $t$ the current does not longer grow because
there is no production of chirality anymore. For large values of
$B_\rmz / E_\rmz$ we find that the current density becomes
approximately constant in time if $\tilde m=0$. If the fermions are
massive the current density exhibits a slowly damped sinusoidal
oscillation for large values of $B_\rmz / E_\rmz$.  The mass also
suppresses the magnitude of the current.

\begin{figure}[t]
\includegraphics[scale=1]{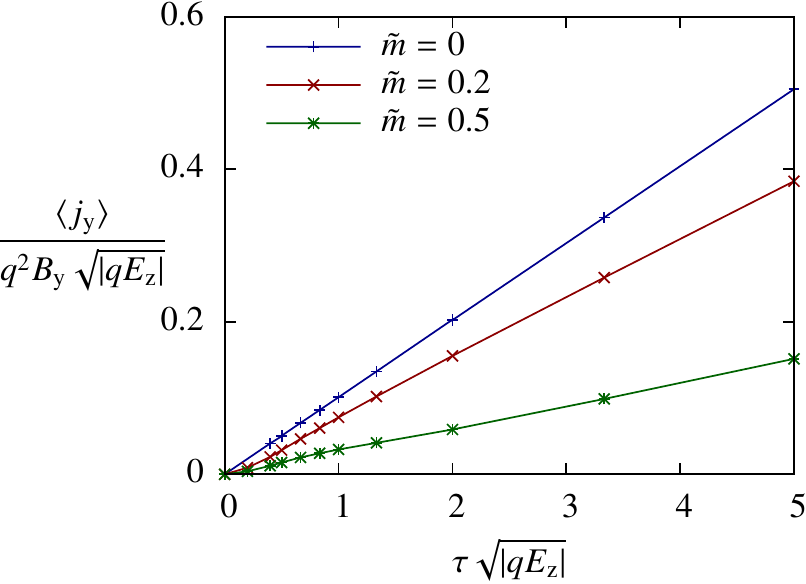}
\caption{Average current density at late times in a pulsed
electric field as a function
of $\tau$ for different $\tilde m$.
\label{fig:jp_sum}
}
\end{figure}

\begin{figure}[t]
\includegraphics[scale=1]{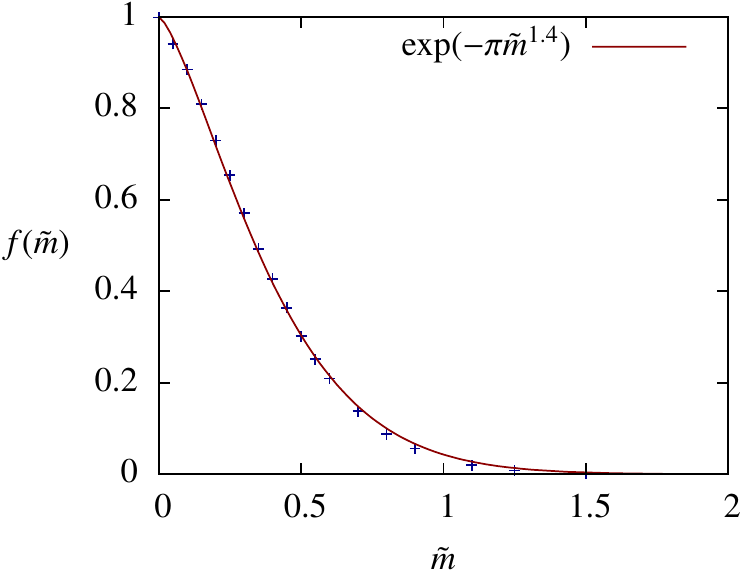}
\caption{The function $f(\tilde m)$ which describes
the mass dependence of the average current density at
late times in a pulsed electric field. Points are numerical
results, solid line is a fit. 
\label{fig:jp_sum_c}
}
\end{figure}

We observe that for smaller values of $B_\rmz/E_\rmz$ the current
density oscillates around the behavior of the current density at large
$B_\rmz/E_\rmz$. Therefore the running average of the current density
seems to be independent of $B_\rmz/E_\rmz$. To investigate the
dependence on $\tau$ we have displayed the running average of the
current density at late times in Fig.~\ref{fig:jp_sum} for
$B_\rmz/E_\rmz = 10$. We find that the running average increases
linear with $\tau$. Through observation of our numerical results we
find that for all values of $B_\rmz/E_\rmz$ the running average of the
current density at late times is summarized by the following formula
\begin{equation} 
  \langle j_\rmy \rangle = \frac{q^2 \vert q \vert \tau}{\pi^2} B_\rmy E_\rmz f(\tilde m) 
 \mathrm{sgn}(B_\mathrm{z}),
\end{equation}
where we have displayed $f(\tilde m)$ for different values of $\tilde
m$ in Fig.~\ref{fig:jp_sum_c}. We find a reasonable fit to our data
with the function $f(\tilde m) = \exp(-\pi \tilde m^{1.4})$.

\section{Conclusions}
\label{sec:conclusions}
In this article we have investigated the real-time dynamics of the
chiral magnetic effect using linear response theory. We have
considered a field configuration in which a homogeneous
(chromo)electric field with arbitrary time-dependence lies parallel to
a homogeneous and static (chromo)magnetic field. These parallel fields
are the source of the chirality. To this field configuration we have
added a perpendicular homogeneous and static magnetic field. We have
computed the induced current density along this perpendicular magnetic
field explicitly for a sudden switch-on and a pulsed electric field.

In the sudden switch-on electric field we have obtained excellent
agreement with an earlier independent analytic computation of the
current density. In the pulsed electric field we could summarize the
induced current density that we have obtained numerically with a
simple analytic formula.

The main purpose of this article was to demonstrate the dynamics of
the chiral magnetic effect using linear response theory. We hope that
our results will be extended to other interesting field configurations
in the future.  For example, in heavy ion collisions it would be
important to answer the question whether there is enough time for the
chiral magnetic effect to occur in the quickly decaying magnetic
field. This question could be addressed using our methodology.  The
chiral magnetic effect could also be investigated with lasers that
create strong electromagnetic fields. For that purpose it would be
important to extend the results to field configurations that are as
close to the experimental situation as possible.

As a side result of our work we have obtained a derivation of the
induced current density in static homogeneous parallel electric and
magnetic fields. We have also obtained an analytic formula for the
chirality production for massive fermions in static homogeneous
parallel electric and magnetic fields.


\section*{Acknowledgments}
I would like to thank Gerald Dunne, Kenji Fukushima,  Dmitri Kharzeev, 
Larry McLerran and Vladimir Skokov for discussions.  The work of H.J.W.\ was
supported by the Extreme Matter Institute (EMMI) and by the Alexander
von Humboldt Foundation.

\appendix
\section{Relations involving the function 
\texorpdfstring{$g_{\vec p s}(\rmx)$}{gps(x)}}
\label{app:rel}
To evaluate these integrals we will make use of the following four
identities which directly follow from the properties of the Hermite
polynomials,
\begin{multline}
\left. 
\int_{-\infty}^{\infty} \rmd \rmx' \rmx' 
g_{\vec p s}(\rmx') g_{\vec p' s}(\rmx') 
\right \vert_{p_\rmy'=p_\rmy}
= \frac{p_\rmy}{qB_\rmz} \delta_{k, k'} 
\left( 
1 - \delta_{k,0} \delta_{s-}
\right)
\\
+ \frac{1}{\sqrt{2 \vert q B_\rmz \vert}}
\left(
\sqrt{k}  \, \delta_{k-s,k'}
+
\sqrt{k+s} \, \delta_{k+s,k'}
\right),
\end{multline}
\begin{multline}
\left. 
\frac{1}{qB_\rmz}\int_{-\infty}^{\infty} \frac{\rmd p_\rmy}{2 \pi}
p_\rmy g_{\vec p-s}( \rmx ) g_{\vec p' s}( \rmx) 
\right \vert_{p_\rmy'=p_\rmy}
= 
\frac{\vert q B_\rmz \vert}{2 \pi}  \Biggl [
\rmx \delta_{k-s, k'}
\\
- 
\frac{1}{\sqrt{2 \vert q B_\rmz \vert}}
\left(
\sqrt{k-s} \, \delta_{k-2s,k'}
+
\sqrt{k}\, \delta_{k,k'}
\right)
\Biggr ],
\end{multline}
\begin{equation}
\left.
\int_{-\infty}^{\infty} \frac{\rmd p_\rmy}{2 \pi}
g_{\vec p-s}( \rmx ) g_{\vec p' s}( \rmx ) 
\right \vert_{p_\rmy'=p_\rmy}
=
\frac{\vert q B_\rmz \vert}{2 \pi}
\delta_{k-s,k'}
.
\end{equation}
\begin{equation}
\left.
\int_{-\infty}^{\infty} \rmd \rmx \,
g_{\vec ps}( \rmx ) g_{\vec p' -s}( \rmx ) 
\right \vert_{p_\rmy'=p_\rmy}
=
\delta_{k+s,k'}
.
\end{equation}

\section{Wave functions in a sudden switch-on electric field}
\label{app:sol}
We will review the explicit solutions for the wave functions
$f^\pm_{\vec p s}(t)$ for a sudden switch-on electric field of the form $E(t) =
E_\rmz \theta(t)$ \cite{N69}. Then we will evaluate an integral that
is necessary for computing the induced current and chirality
production.

For the sudden switch-on electric field we have $A_{\rmz}(t) = -E_\rmz t
\theta(t)$.  For $t<0$ we can use the wave functions in vanishing
electromagnetic field, given in Eq.~(\ref{eq:fpmfree}).  If $t>0$ the
electric field is not longer vanishing. It follows from combining
Eqs.~(\ref{eq:1orderelA}) and (\ref{eq:1orderelB}) that $f^\pm_{\vec p s}(t)$
then satisfies
\begin{equation}
 \left[ - \partial_t^2 - q^2 E_\rmz^2 
\left(t + \frac{p_\rmz}{q E_\rmz} \right)^2  + \mathrm{i} s q E_\rmz 
\right]
f^\pm_{\vec p s}(t) = \kappa^2  f^\pm_{\vec p s}(t).
\label{eq:harmoel}
\end{equation}
Eq.~(\ref{eq:harmoel}) is an eigenvalue equation for a particle in an
upside-down harmonic potential. There are no bound states, so $\kappa$
is not quantized.  The solution is a linear combination of parabolic
cylinder functions $D_\nu(z)$,
\begin{equation}
f_{\vec p s}^\pm(t) = \alpha^\pm_s 
D_{\nu_s}( \xi ) 
+ 
\beta^\pm_s
D_{\nu_s}( -\xi),
\label{eq:elsoltp}
\end{equation}
where $\nu_s = -(s \, \mathrm{sgn}(qE_\rmz) + 1 + \rmi \kappa^2 / \vert q E_\rmz
\vert)/2$ and $\xi = \sqrt{2 \vert q E_\rmz \vert}\rme^{\rmi \pi /4}
(t + p_\rmz / qE_\rmz)$.  

In the conventional normalization the parabolic cylinder functions
are given explicitly by the following integrals
\begin{equation}
 D_\nu(z) = 
\frac{1}{\Gamma(-\nu)} \rme^{-\tfrac{1}{4}z^2 }
\int_0^{\infty} \rmd t \, t^{-\nu -1} 
\rme^{-z t - \tfrac{1}{2} t^2 },
\end{equation}
for $\mathrm{Re}(\nu) < 0$, and
\begin{equation}
D_{\nu}(z) =
\sqrt{\frac{2}{\pi}} \rme^{\tfrac{1}{4}z^2 }
\int_0^{\infty} \rmd t \, t^{\nu} 
\cos\left(zt - \tfrac{1}{2} \pi \nu \right)
\rme^{- \tfrac{1}{2} t^2}
\label{eq:asympara}
\end{equation}
for $\mathrm{Re}(\nu) > -1$.
Using these relations one can show that $D_\nu(0) = 2^{\nu/2}
\sqrt{\pi} / \Gamma[(1-\nu)/2]$.  Asymptotically the parabolic
cylinder functions behave as follows
\begin{equation}
\lim_{\vert z \vert \rightarrow \infty} D_\nu(z) = 
z^\nu \rme^{-\tfrac{1}{4} z^2} 
- c_\nu \frac{\sqrt{2\pi}}{\Gamma(-\nu)} 
z^{-\nu - 1} \rme^{\tfrac{1}{4} z^2},
\end{equation}
with $c_\nu = 0$ for $\vert \mathrm{arg}(z) \vert < \frac{3 \pi}{4}$,
$c_\nu= \exp(\rmi \pi \nu)$ for $\frac{\pi}{4} < \vert \mathrm{arg}(z)
\vert < \frac{5 \pi}{4}$, and for $c_\nu = \exp(-\rmi\pi\nu)$ for
$-\frac{5\pi}{4} < \vert \mathrm{arg}(z) \vert < -\frac{\pi}{4}$.
These asymptotic functions describe the real parabolic cylinder
functions very good if $\vert z \vert > 2 \vert \nu \vert$.

The constants $\alpha^\pm_s$ and $\beta^\pm_s$ of
Eq.~(\ref{eq:elsoltp}) can be found by requiring continuity of
$f^\pm_{\vec p s}(t)$ and its derivative at $t=0$. This gives the following two
equations
\begin{eqnarray}
\alpha^\pm_s &=&
\frac{1}{W} 
\sqrt{\frac{p_0 \mp s p_\rmz}{p_0}}
\Bigl [
-\rme^{\rmi \pi/4} \sqrt{2\vert q E_\rmz \vert}
D_{\nu_s}'( -\xi_0) 
\nonumber \\
&& \quad \quad \quad \quad \quad \quad \quad
\pm \rmi p_0
D_{\nu_s}( -\xi_0)
\Bigr ],
\label{eq:alphas1}
\\
\beta_s^\pm &=&
\frac{1}{W} 
\sqrt{\frac{p_0 \mp s p_\rmz}{p_0}}
\Bigl [
-\rme^{\rmi \pi/4} \sqrt{2 \vert q E_\rmz \vert}
D_{\nu_s}'( \xi_0) 
\nonumber 
\\
&& \quad \quad \quad \quad \quad \quad \quad
\mp  \rmi p_0
D_{\nu_s}(\xi_0)
\Bigr ],
\label{eq:betas1}
\end{eqnarray}
where $\xi_0 = \sqrt{2} \rme^{\rmi \pi/4} p_\rmz / \vert q E_\rmz \vert^{1/2} \,
\mathrm{sgn}(q E_\rmz) $ and $W$ denotes the Wronskian of the two
independent solutions presented in Eq.~(\ref{eq:elsoltp}). Applying
Abel's differential equation identity to Eq.~(\ref{eq:harmoel}) shows
that the Wronskian is independent of $\xi$. Hence without loss of
generality we can evaluate $W$ at $\xi = 0$, which yields $W = 2
\rme^{\rmi \pi/4} \sqrt{\pi \vert q E_\rmz \vert} / \Gamma(-\nu_s)$.
Using the relation $D_\nu'(z) = \tfrac{1}{2} z D_\nu(z) - D_{\nu+1}(z)$ 
we can simplify Eqs.~(\ref{eq:alphas1}) and (\ref{eq:betas1}) to
\begin{eqnarray}
\alpha^\pm_s &=&
\frac{\Gamma(-\nu_s)}{\sqrt{2 \pi} }
\sqrt{\frac{p_0 \mp s p_\rmz}{p_0}}
\Bigl [
D_{\nu_s+1}( -\xi_0)
\nonumber 
\\
&&
+ 
\frac{\rme^{\rmi \pi/4}}
{\sqrt{2 \vert q E_\rmz \vert}}
\left( p_\rmz \mathrm{sgn}(qE_\rmz) \pm p_0 \right)
D_{\nu_s}( -\xi_0)
\Bigr],
\\
\beta_s^\pm &=&
\frac{\Gamma(-\nu_s)}{\sqrt{2 \pi} }
\sqrt{\frac{p_0 \mp s p_\rmz}{p_0}}
\Bigl[
D_{\nu_s+1}( \xi_0)
\nonumber 
\\
&& - 
\frac{\rme^{\rmi \pi/4}}
{\sqrt{2 \vert q E_\rmz \vert}}
\left( p_\rmz \mathrm{sgn}(qE_\rmz)
\pm p_0 \right)
D_{\nu_s}(\xi_0)
\Bigr ].
\end{eqnarray}

Using the asymptotic expansion of the parabolic cylinder functions,
Eq.~(\ref{eq:asympara}), it is possible to obtain a very good
approximation for the function $f_{\vec p s}^\pm(t)$ away from the points $p_\rmz
= -qE_\rmz t$ and $p_\rmz = 0$. Let us now for a moment choose $qE_\rmz> 0$ and
take $\sqrt{\vert q E_\rmz \vert} t \gg 1$.  In order to obtain the induced
current we need to evaluate $\vert f_{\vec p+}^+(t) \vert^2$, for which we
find after taking the dominating terms in the asymptotic expansion and
approximating $p_0$ by $\vert p_\rmz\vert$ the following
\begin{equation}
 \vert f_{\vec p +}^+(t)\vert^2 \approx
\left \{
\begin{array}{ll}
2
&
\mathrm{for}\; p_\rmz \lesssim -qE_\rmz t - \Delta,
\\
2 \rme^{-\frac{\pi \kappa^2}{\vert q E_\rmz \vert}} + g(\zeta)
&
\mathrm{for}\; -qE_\rmz t + \Delta
\lesssim  p_\rmz \lesssim 
-\Delta,
\\
0
&
\mathrm{for}\; p_\rmz \gtrsim \Delta.
\end{array}
\right.
\label{eq:absfpsqapprox}
\end{equation}
where $\Delta = 2 \vert q E_\rmz \vert^{1/2}
\vert 1 + \frac{\rmi \kappa^2}{2\vert q E_\rmz \vert} \vert$,
$\zeta = \sqrt{\vert qE_\rmz \vert}(t + p_\rmz/qE_\rmz)$, and
\begin{multline}
g(\zeta) = 
\frac{2}{\sqrt{\pi}} 
\frac{1}{\zeta}
\left[ 
\rme^{-\frac{\pi \kappa^2}{4\vert q E_\rmz \vert}}
-
\rme^{-\frac{5 \pi \kappa^2}{4\vert q E_\rmz \vert}}
\right] 
\\
\times
\mathrm{Re}
\left[
\Gamma
\left(1 - \tfrac{\rmi \kappa^2}{2 \vert q E_\rmz \vert}
\right)
\rme^{\rmi \zeta^2
+
\frac{\rmi \kappa^2}{\vert q E_\rmz \vert}
\log \vert \sqrt{2} \zeta \vert + \rmi \pi / 4
}
\right].
\end{multline}

As follows from Eq.~(\ref{eq:regj3}) we have to evaluate the
following integral to obtain the induced current
\begin{equation}
I(\kappa, t) = 
\lim_{\epsilon \rightarrow 0}
\int_{-\infty}^{\infty} \frac{\rmd p_\rmz}{2 \pi}
\rme^{- \epsilon (p_\rmz + q E_\rmz t)^2} \left[ \vert f_{\vec p +}^+(t) \vert^2
- 1 \right].
\end{equation}
In general this integral can only be evaluated numerically.  But using
Eq.~(\ref{eq:absfpsqapprox}) we can obtain $I(\kappa, t)$ exactly for
large $t$. Firstly one realizes that taking the limit $\epsilon
\rightarrow 0$ implies that we have to integrate $\vert f_{\vec p +}^+(t)
\vert^2 - 1$ over $p_\rmz$ in an interval symmetric around $p_\rmz = -
q E_\rmz t$. The contribution to the integral in the region where the
approximation Eq.~(\ref{eq:absfpsqapprox}) breaks down can be bounded
from below and above by a time-independent constant.  The
contributions for $p_\rmz < - 2 q E_\rmz t$ and $p_\rmz > 0$ will
cancel against each other. The non-vanishing contribution comes from
the intermediate region in which $-2 qE_\rmz t < p_\rmz < 0$. The
integral over the rapidly oscillating function $g(\zeta)$ is
subdominant in the large $t$ limit. The dominant contribution in the
large $t$ limit is $q E_\rmz t$ coming from the interval $-2qE_\rmz t
< p_\rmz < - qE_\rmz t$ and $qE_\rmz t[2 \exp(-\pi \kappa^2/ \vert q
E_\rmz \vert) - 1]$ from the interval $-qE_\rmz t < p_\rmz<0$.  Ading
both contributions yields
\begin{equation}
\lim_{t\rightarrow \infty} \frac{1}{t} I(\kappa, t) 
= \frac{q E_\rmz }{\pi}
\rme^{-  \frac{\pi \kappa^2}{\vert q E_\rmz \vert} }
.
\label{eq:ilimit}
\end{equation}
By repeating the analysis for $q E_\rmz < 0$ one can verify that the
sign of Eq.~(\ref{eq:ilimit}) is correct, so Eq.~(\ref{eq:ilimit})
holds for all values of $q E_\rmz$.

$\phantom{nothing}$



\begin{thebibliography}{99}


\bibitem{S51}
J.~S.~Schwinger,
\href{http://dx.doi.org/PhysRev.82.664}{Phys.\ Rev.\ {\bf 82} (1951) 664}.

\bibitem{ABJ}
 S.~L.~Adler,
\href{http://dx.doi.org/10.1103/PhysRev.177.2426}{Phys.\ Rev.\  {\bf 177}, 2426 (1969)}; 
J.~S.~Bell and R.~Jackiw,
\href{http://dx.doi.org/10.1007/BF02823296}{Nuovo\ Cim.\  {\bf A60},~47~(1969)}.


\bibitem{V80} 
A.~Vilenkin,
\href{http://dx.doi.org/10.1103/PhysRevD.22.3080}{Phys.\ Rev.\ D {\bf 22}, 3080 (1980)}.


\bibitem{KMW}
  D.~E.~Kharzeev, L.~D.~McLerran and H.~J.~Warringa,
\href{http://dx.doi.org/10.1016/j.nuclphysa.2008.02.298}{Nucl.\ Phys.\  A {\bf 803}, 227 (2008)}.





\bibitem{mu5}
  A.~Y.~Alekseev, V.~V.~Cheianov and J.~Frohlich,
\href{http://dx.doi.org/10.1103/PhysRevLett.81.3503}{Phys.\ Rev.\ Lett.\ {\bf 81}, 3503 (1998)};
  K.~Fukushima, D.~E.~Kharzeev and H.~J.~Warringa,
\href{http://dx.doi.org/10.1103/PhysRevD.78.074033}{Phys.\ Rev.\  D {\bf 78}, 074033 (2008)};
 D.~E.~Kharzeev and H.~J.~Warringa,
\href{http://dx.doi.org/10.1103/PhysRevD.80.034028}{Phys.\ Rev.\  D {\bf 80}, 034028 (2009)};
  H.~-U.~Yee,
\href{http://dx.doi.org}{JHEP {\bf 0911}, 085 (2009)};
  A.~Rebhan, A.~Schmitt and S.~A.~Stricker,
\href{http://dx.doi.org/10.1007/JHEP01(2010)026}{JHEP {\bf 1001}, 026 (2010)};
  V.~A.~Rubakov,
\href{http://www.arxiv.org/abs/1005.1888}{arXiv:1005.1888} [hep-ph];
  A.~Gynther, K.~Landsteiner, F.~Pena-Benitez and A.~Rebhan,
 \href{http://dx.doi.org/10.1007/JHEP02(2011)110}{JHEP {\bf 1102}, 110 (2011)};
V.~D.~Orlovsky and V.~I.~Shevchenko,
\href{http://dx.doi.org/10.1103/PhysRevD.82.094032}{Phys.\ Rev.\ D {\bf 82}, 094032 (2010)};
  L.~Brits and J.~Charbonneau,
\href{http://dx.doi.org/10.1103/PhysRevD.83.126013}{Phys.\ Rev.\ D {\bf 83}, 126013 (2011)};
D.~Hou, H.~Liu and H.~-c.~Ren,
\href{http://dx.doi.org/10.1007/JHEP05(2011)046}{JHEP {\bf 1105}, 046 (2011)};
C.~Hoyos, T.~Nishioka and A.~O'Bannon,
\href{http://dx.doi.org/10.1007/JHEP10(2011)084}{JHEP {\bf 1110}, 084 (2011)};
A.~Yamamoto,
\href{http://dx.doi.org/10.1103/PhysRevD.84.114504}{Phys.\ Rev.\ D {\bf 84}, 114504 (2011)};
D.~T.~Son and N.~Yamamoto,
\href{http://www.arxiv.org/abs/1203.2697}{arXiv:1203.2697} [cond-mat.mes-hall].

\bibitem{mu5f}
  K.~Fukushima, D.~E.~Kharzeev and H.~J.~Warringa,
\href{http://dx.doi.org/10.1016/j.nuclphysa.2010.02.003}{Nucl.\ Phys.\ A {\bf 836}, 311 (2010)}.

\bibitem{lattice}
  P.~V.~Buividovich, M.~N.~Chernodub, E.~V.~Luschevskaya and M.~I.~Polikarpov,
\href{http://dx.doi.org/10.1103/PhysRevD.80.054503}{Phys.\ Rev.\  D {\bf 80}, 054503 (2009)};
  M.~Abramczyk, T.~Blum, G.~Petropoulos and R.~Zhou,
\href{http://www.arxiv.org/abs/0911.1348}{arXiv:0911.1348} [hep-lat].


\bibitem{NAM09}  
S.~i.~Nam,
\href{http://dx.doi.org/10.1103/PhysRevD.80.114025}{Phys.\ Rev.\  D {\bf 80}, 114025 (2009)};
S.~i.~Nam,
\href{http://dx.doi.org/10.1103/PhysRevD.82.045017}{Phys.\ Rev.\ D {\bf 82}, 045017 (2010)}.


\bibitem{BDK12} 
G.~Basar, G.~V.~Dunne and D.~E.~Kharzeev,
\href{http://dx.doi.org/10.1103/PhysRevD.85.045026}{Phys.\ Rev.\ D {\bf 85}, 045026 (2012)}.

\bibitem{FKW10} 
K.~Fukushima, D.~E.~Kharzeev and H.~J.~Warringa,
\href{http://dx.doi.org/10.1103/PhysRevLett.104.212001}{Phys.\ Rev.\ Lett.\  {\bf 104}, 212001 (2010)}.


\bibitem{D08} 
H.~Gies,
\href{http://dx.doi.org/10.1140/epjd/e2009-00006-0}{Eur.\ Phys.\ J.\ D {\bf 55}, 311 (2009)};
M.~Marklund and J.~Lundin,
\href{http://dx.doi.org/10.1140/epjd/e2009-00169-6}{Eur.\ Phys.\ J.\ D {\bf 55}, 319 (2009)};
G.~V.~Dunne,
\href{http://dx.doi.org/10.1140/epjd/e2009-00022-0}{Eur.\ Phys.\ J.\ D {\bf 55}, 327 (2009)}.



\bibitem{glasma}
D.~Kharzeev, A.~Krasnitz and R.~Venugopalan,
\href{http://dx.doi.org/10.1016/S0370-2693(02)02630-8}{Phys.\ Lett.\ B {\bf 545}, 298 (2002)};
T.~Lappi and L.~McLerran,
\href{http://dx.doi.org/10.1016/j.nuclphysa.2006.04.001}{Nucl.\ Phys.\  A {\bf 772}, 200 (2006)};


\bibitem{SIT}
  H.~Minakata and B.~Muller,
\href{http://dx.doi.org/10.1016/0370-2693(96)00337-1}{Phys.\ Lett.\ B {\bf 377}, 135 (1996)};
  V.~Skokov, A.~Y.~Illarionov and V.~Toneev,
\href{http://dx.doi.org/10.1142/S0217751X09047570}{Int.\ J.\ Mod.\ Phys.\  A {\bf 24}, 5925 (2009)};
\  V.~Voronyuk, V.~D.~Toneev, W.~Cassing, E.~L.~Bratkovskaya, V.~P.~Konchakovski and S.~A.~Voloshin,
\href{http://dx.doi.org/10.1103/PhysRevC.83.054911}{Phys.\ Rev.\ C {\bf 83}, 054911 (2011)};
  W.~-T.~Deng and X.~-G.~Huang,
\href{http://dx.doi.org/10.1103/PhysRevC.85.044907}{Phys.\ Rev.\ C {\bf 85}, 044907 (2012)}.


\bibitem{M11} 
  G.~D.~Moore and M.~Tassler,
\href{http://dx.doi.org/10.1007/JHEP02(2011)105}{JHEP {\bf 1102}, 105 (2011)}.

\bibitem{KKZ}
  D.~Kharzeev,
\href{http://dx.doi.org/10.1016/j.physletb.2005.11.075}{Phys.\ Lett.\ B {\bf 633}, 260 (2006)}; 
  D.~Kharzeev and A.~Zhitnitsky,
\href{http://dx.doi.org/10.1016/j.nuclphysa.2007.10.001}{Nucl.\ Phys.\  A {\bf 797}, 67 (2007)}.


\bibitem{V04}
  S.~A.~Voloshin,
\href{http://dx.doi.org/10.1103/PhysRevC.70.057901}{Phys.\ Rev.\  C {\bf 70}, 057901 (2004)}.

\bibitem{STAR}
  B.~I.~Abelev {\it et al.}  [STAR Collaboration],
\href{http://dx.doi.org/10.1103/PhysRevLett.103.251601}{Phys.\ Rev.\ Lett.\ {\bf 103}, 251601 (2009)};
  B.~I.~Abelev {\it et al.}  [STAR Collaboration],
\href{http://dx.doi.org/10.1103/PhysRevC.81.054908}{Phys.\ Rev.\ C {\bf 81}, 054908 (2010)}.

\bibitem{ALICE} 
P.~Christakoglou,
\href{http://dx.doi.org/10.1088/0954-3899/38/12/124165}{J.\ Phys.\ G {\bf 38}, 124165 (2011)}.


\bibitem{alternative} 
  A.~Bzdak, V.~Koch and J.~Liao,
\href{http://dx.doi.org/10.1103/PhysRevC.81.031901}{Phys.\ Rev.\ C {\bf 81}, 031901 (2010)};
  M.~Asakawa, A.~Majumder and B.~Muller,
\href{http://dx.doi.org/10.1103/PhysRevC.81.064912}{Phys.\ Rev.\ C {\bf 81}, 064912 (2010)};
  J.~Liao, V.~Koch and A.~Bzdak,
\href{http://dx.doi.org/10.1103/PhysRevC.82.054902}{Phys.\ Rev.\ C {\bf 82}, 054902 (2010)};
  H.~Petersen, T.~Renk and S.~A.~Bass,
\href{http://dx.doi.org/10.1103/PhysRevC.83.014916}{Phys.\ Rev.\ C {\bf 83}, 014916 (2011)};
  A.~Bzdak, V.~Koch and J.~Liao,
\href{http://dx.doi.org/10.1103/PhysRevC.83.014905}{Phys.\ Rev.\ C {\bf 83}, 014905 (2011)};
B.~Muller and A.~Schafer,
\href{http://dx.doi.org/10.1103/PhysRevC.82.057902}{Phys.\ Rev.\ C {\bf 82}, 057902 (2010)}.



\bibitem{R72} 
V.~I.~Ritus,
\href{http://dx.doi.org/10.1016/0003-4916(72)90191-1}{Annals Phys.\  {\bf 69}, 555 (1972)};
V.~I.~Ritus,
Sov.\ Phys.\ JETP {\bf 48}, 788 (1978)
[Zh.\ Eksp.\ Teor.\ Fiz.\  {\bf 75}, 1560 (1978)].

\bibitem{PeskinSchroeder}
M.~E.~Peskin and D.~V.~Schroeder.
{\it An introduction to Quantum Field Theory},
(Perseus Books 1995).



\bibitem{N69}
A.~I.~Nikishov,
Zh.\ Eksp.\ Teor.\ Fiz.\  {\bf 57} (1969) 1210.

\bibitem{NN70} 
  N.~B.~Narozhnyi and A.~I.~Nikishov,
  Yad.\ Fiz.\  {\bf 11}, 1072 (1970)
  [Sov.\ J.\ Nucl.\ Phys.\  {\bf 11}, 596 (1970)].


\bibitem{AGP83} 
J.~Ambjorn, J.~Greensite and C.~Peterson,
\href{http://dx.doi.org/10.1016/0550-3213(83)90585-0}{Nucl.\ Phys.\ B {\bf 221}, 381 (1983)}.

\bibitem{K92} 
Y.~Kluger, J.~M.~Eisenberg, B.~Svetitsky, F.~Cooper and E.~Mottola,
\href{http://dx.doi.org/10.1103/PhysRevD.45.4659}{Phys.\ Rev.\ D {\bf 45}, 4659
(1992)}.

\bibitem{PhotonPol} 
L.~F.~Urrutia,
\href{http://dx.doi.org/10.1103/PhysRevD.17.1977}{Phys.\ Rev.\ D {\bf 17}, 1977 (1978)};
 V.~P.~Barashev, A.~E.~Shabad and S.~M.~Shvartsman,
  Sov.\ J.\ Nucl.\ Phys.\  {\bf 43}, 617 (1986)
  [Yad.\ Fiz.\  {\bf 43}, 964 (1986)];
  C.~Schubert,
\href{http://dx.doi.org/10.1016/S0550-3213(00)00423-5}{Nucl.\ Phys.\ B {\bf 585}, 407 (2000)};


\bibitem{PhotonPolMag}
S.~L.~Adler,
\href{http://dx.doi.org/10.1016/0003-4916(71)90154-0}{Annals Phys.\  {\bf 67} (1971) 599};
W.~-y.~Tsai,
\href{http://dx.doi.org/10.1103/PhysRevD.10.2699}{Phys.\ Rev.\ D {\bf 10}, 2699 (1974)};
A.~E.~Shabad,
\href{http://dx.doi.org/10.1016/0003-4916(75)90144-X}{Annals Phys.\  {\bf 90}, 166 (1975)}.

\bibitem{PhotonPolMagRecent}
K.~Fukushima,
\href{http:///dx.doi.org/10.1103/PhysRevD.83.111501}{Phys.\ Rev.\ D
  {\bf 83}, 111501 (2011)};
E.~J.~Ferrer, V.~de la Incera and A.~Sanchez,
\href{http://dx.doi.org/10.1103/PhysRevLett.107.041602}{Phys.\ Rev.\
  Lett.\ {\bf 107}, 041602 (2011)};
H.~Gies and L.~Roessler, 
\href{http://dx.doi.org/10.1103/PhysRevD.84.065035}{Phys.\ Rev.\ D {\bf 84}, 065035 (2011)};
 F.~Karbstein, L.~Roessler, B.~Dobrich and H.~Gies,
 \href{http://www.arxiv.org/abs/1111.5984}{arXiv:1111.5984} [hep-ph];
 N.~Sadooghi and F.~Taghinavaz,
\href{http://www.arxiv.org/abs/1203.5634}{arXiv:1203.5634} [hep-ph].



\bibitem{GG08} 
S.~P.~Gavrilov and D.~M.~Gitman,
\href{http://dx.doi.org/10.1103/PhysRevD.78.045017}{Phys.\ Rev.\ D {\bf 78}, 045017 (2008)}.

\bibitem{T09} 
N.~Tanji,
\href{http://dx.doi.org/10.1016/j.aop.2009.03.012}{Annals Phys.\ {\bf 324}, 1691 (2009)}.


\bibitem{BT70} 
F.~V.~Bunkin and I.~I.~Tugov, 
Sov.\ Phys.\ Dokl.\ {\bf 14} (1970) 678.

\bibitem{D04}
G.~V.~Dunne,
\href{http://www.arxiv.org/abs/hep-th/0406216}{arXiv:hep-th/0406216}.

\bibitem{KP06}
S.~P.~Kim and D.~N.~Page,
\href{http://dx.doi.org/10.1103/PhysRevD.73.065020}{Phys.\ Rev.\  D {\bf 73} (2006) 065020}.  

\bibitem{CM08}
T.~D.~Cohen and D.~A.~McGady,
\href{http://dx.doi.org/10.1103/PhysRevD.78.036008}{Phys.\ Rev.\  D {\bf 78} (2008) 036008}.


















\end{thebibliography}
\end{document}